\documentclass[11pt]{amsart}
\usepackage{etex}
\newtheorem{Theorem}{Theorem}[section]
\newtheorem{Proposition}[Theorem]{Proposition}
\newtheorem{Lemma}[Theorem]{Lemma}

\newtheorem{Corollary}[Theorem]{Corollary}

\newtheorem{Assumption 2}[Theorem]{Assumption 2}

\numberwithin{equation}{section}

\usepackage{amsmath}
\usepackage{amssymb}
\usepackage{lmodern}
\usepackage{stmaryrd}
\usepackage{ulem}
\usepackage{yfonts}
\usepackage{amsbsy}
\usepackage{pst-xkey}
\usepackage{pst-all,pst-gr3d}
\usepackage{gensymb}
\usepackage{tikz}

\def\k#1{\kern#1em}
\def\Ib#1{{I\kern-.25em#1}}
\def\Ibb#1{{I\kern-.23em#1}}

\def\AA{{\mathbb A}}
\def\BB{\mathbb B}
\def\CC{{\mathbb C}}

\def\RR{{\mathbb{R}}}

\def\vci{\vrule  width.02em height1.47ex depth-.0ex}
\def\11{{\rm\k{.2}\vci\k{-.37}1}}

\def\fin{{\begin{flushright}
\it{Q.E.D.}
\end{flushright}}}

\begin{document}

\address{Universit\'e de Bordeaux, Institut de Math\'ematiques, UMR CNRS 5251, F-33405 Talence Cedex}

\email{alain.bachelot@u-bordeaux.fr}

\title{On the Klein-Gordon equation near a De Sitter Brane}

\author{Alain BACHELOT}

\begin{abstract}
In this paper we investigate the Klein-Gordon equation in the past causal domain of a De Sitter brane imbedded in an Anti-de Sitter bulk.
We solve the global mixed hyperbolic problem. We prove that any finite energy solution can be expressed as a Kaluza-Klein tower that is a superposition of free fields in the Steady State Universe, of which we study the asymptotic behaviours. We show that the leading term of a gravitational fluctuation is a massless graviton, i.e. the De Sitter brane is linearly stable. Beyond the Cauchy horizon, the energy of the waves can tend to infinity at the moment the brane hits the conformal time-like boundary.
\end{abstract}

\maketitle


\pagestyle{myheadings}
\markboth{\centerline{\sc Alain Bachelot}}{\centerline{\sc {Klein-Gordon equation near a De Sitter Brane}}}

\section{Introduction}

The Klein-Gordon equation with mass $M\geq 0$ on a Lorentzian manifold $({\mathcal M},g)$ is defined by
\begin{equation}
\square_{\mathcal M}u+M^2u=0,\;\;\square_{\mathcal M}:=\frac{1}{\sqrt{\mid g\mid}}\frac{\partial}{\partial x^{\mu}}\left(\sqrt{\mid g\mid}g^{\mu \nu}\frac{\partial}{\partial x^{\nu}}\right),\;\;\mid g\mid:=\left\vert det\left(g_{\mu\nu}\right)\right\vert.
 \label{eq}
\end{equation}
In this work, we consider the case where ${\mathcal M}$ is a bulk in the $(1+4)$-dimensional Anti-de Sitter space $AdS^5$, and the time-like part ${\mathcal B}$ of its boundary $\partial{\mathcal M}$  is a $(1+3)$-dimensional De Sitter Brane on which we impose a Robin type condition (or the Dirichlet condition) for $u$:
\begin{equation}
n^{\mu}\frac{\partial u}{\partial x^{\mu}}+cu=0,\;\; c\in\RR,\;\; (or\;\;u=0,\;\; if \;\; c=\infty),
 \label{cl}
\end{equation}
where $n^{\mu}$ is the outgoing unit normal vector at ${\mathcal B}$ and $c$ is a parameter characterizing the strenght of the coupling of the field with the brane. We investigate these waves near the brane in the sense that we assume that $\mathcal{M}$ is the past causal set of the brane $\mathcal B$. This model is very important in String Cosmology and has been deeply investigated by the physicists, {\it e.g.} in \cite{gorbunov}, \cite{langlois1}, \cite{langlois2}, \cite{mannheim}, \cite{parikh}. Nevertheless a rigorous mathematical investigation was missing, specially as regards the functional framework, the global mixed hyperbolic problem and the spectral analysis of the hamiltonian, which are the purposes of this paper.

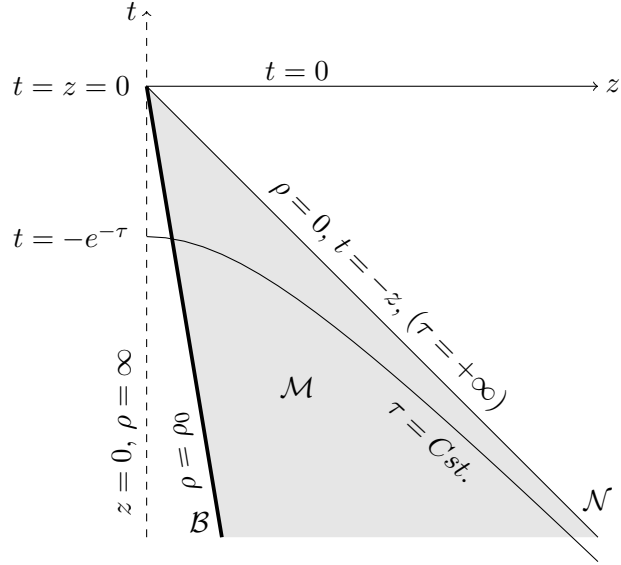
\begin{figure}
\begin{tikzpicture}
\draw  [->] [dashed] (0,-2) -- (0,5) node[pos=0.2,sloped,above] {$z=0$, $\rho=\infty$};
\draw [->] (0,4) -- (6,4);
\fill[color=gray!20]
(0,4) -- (6,-2)
-- (1,-2)--(0,4)
-- cycle;
\draw (6.2,4) node {$z$};
\draw (-0.2,5) node {$t$};
\draw (-1,4) node{$t=z=0$};
\draw (0.7,-1.8) node{$\mathcal B$};
\draw (2,0) node{$\mathcal M$};
\draw (0,4) -- (6,-2) node[midway,sloped,above] {$\rho=0$, $t=-z$, $(\tau=+\infty)$};
\draw (6,-1.5) node{$\mathcal N$};
\draw (2,4.2) node{$t=0$};
\draw [line width=0.5mm](0,4) -- (1,-2) node[pos=0.8,sloped,above, rotate=180] {$\rho=\rho_0$};
\draw [domain=0:6] plot (\x, {4-sqrt((\x)^2+4)});
\draw (-1,2) node{$t=-e^{-\tau}$};
\draw (3.8,-0.7) node[rotate=-43] {$\tau=Cst.$};

\end{tikzpicture}
\caption{Each point of the picture is $\RR^3_{\mathbf x}$. The manifold $\mathcal M$ is the grey sector $0<\rho<\rho_0$ of the Poincar\'e patch. The De Sitter brane $\mathcal B$ and the light-like boundary $\mathcal N$ are respectively located at $\rho=\rho_0$ and $\rho=0$. The submanifold $\tau=Cst.$ is a Cauchy hypersurface of which the Cauchy horizon is  $\mathcal N$.}
\label{croba}
\end{figure}

We now describe our geometrical setting (see Figure \ref{croba}). We consider the Poincar\'e patch of the Anti-de Sitter space-time $AdS^5$, that is the non-globally hyperbolic lorentzian manifold
\begin{equation*}
\mathcal{P}:=\RR_t\times\RR^3_{\mathbf x}\times]0,\infty[_z,\;\;ds^2_{AdS}=\left(\frac{1}{z}\right)^2\left(dt^2-d\mathbf{x}^2-dz^2\right).
 \label{}
\end{equation*}
To introduce a De Sitter brane, we fix $\alpha\in]-1,0[$, and we define
\begin{equation*}
{\mathcal B}:=\left\{(t,\mathbf{x},z)\in \mathcal{P};\;\;z=\alpha t\right\}.
 \label{}
\end{equation*}
If we use the following coordinates on ${\mathcal B}$,
\begin{equation*}
T:=\frac{\sqrt{1-\alpha^2}}{\alpha}\log\mid t\mid,\;\;\mathbf{X}:=\alpha^{-1}\mathbf{x},
 \label{}
\end{equation*}
we easily check that the induced metric on $\mathcal{B}$ is given by
\begin{equation*}
ds^2_{\mathcal B}=dT^2-e^{-\frac{2\alpha}{\sqrt{1-\alpha^2}}T}d\mathbf{X}^2,
 \label{}
\end{equation*}
hence we can see that  ${\mathcal B}$ is a De Sitter manifold with scalar curvature $R=12\frac{\alpha^2}{1-\alpha^2}$ (therefore the Hubble constant is $\sqrt{3}\frac{\mid \alpha\mid}{\sqrt{1-\alpha^2}}$). 
More precisely, ${\mathcal B}$ is half of the De Sitter space-time, and it is just the Steady State Universe proposed by Bondi, Gold and Hoyle (see \cite{haw}, p.125). Now the causal past set of ${\mathcal B}$ is given by
\begin{equation*}
\mathcal{O}:=\left\{(t,\mathbf{x},z)\in {\mathcal P};\;\;t<-z<0\right\},
 \label{O}
\end{equation*}
and if we consider ${\mathcal B}$ as a brane with a positive tension, we restrict our study to
\begin{equation*}
{\mathcal M}:=\left\{(t,\mathbf{x},z)\in {\mathcal P};\;\;-\alpha t< z< -t\right\},
 \label{}
\end{equation*}
of which the boundary $\partial{\mathcal M}$ is composed of the time-like part $\mathcal B$ and the light-like submanifold
\begin{equation*}
{\mathcal N}:=\left\{(t,\mathbf{x},z)\in {\mathcal P};\;z=-t\right\}.
 \label{}
\end{equation*}
The unit outgoing normal vector at a point $(t,\mathbf{x},\alpha t)$ is given by $-\frac{\alpha t}{\sqrt{1-\alpha^2}}(\alpha\partial_t+\partial_z)$.
In the $(t,\mathbf{x},z)$ coordinates, the dynamics (\ref{eq}), (\ref{cl}) has the form
\begin{equation*}
\left[\frac{\partial^2}{\partial t^2}-\Delta_{\mathbf x}-\frac{\partial^2}{\partial z^2}+\frac{3}{z}\frac{\partial}{\partial z}+\frac{M^2}{z^2}\right]u=0,\;\;t<0,\;\;\alpha t<z<-t,\;\;{\mathbf x}\in\RR,
 \label{}
\end{equation*}
\begin{equation*}
\alpha\frac{\partial u}{\partial t}+\frac{\partial u}{\partial z}-\frac{c\sqrt{1-\alpha^2}}{\alpha t}u=0\;\;(or\; u=0\;if\; c=\infty),\;\;t<0,\;\;{\mathbf x}\in \RR^3,\;\;z=\alpha t.
 \label{}
\end{equation*}

To avoid the time dependence of the domain of study and of the boundary condition, we introduce new coordinates $(\tau,\rho)$ on the domain $\mathcal O$ (see Figure \ref{croba}) :
\begin{equation*}
\tau:=-\frac{1}{2}\log\left(t^2-z^2\right)\in\RR,\;\;\rho:=\log\left(\sqrt{\frac{t^2}{z^2}-1}-\frac{t}{z}\right)\in[0,\infty[,
 \label{}
\end{equation*}
that is equivalent to
\begin{equation*}
t=-\frac{\cosh(\rho)}{\sinh(\rho)}e^{-\tau}<0,\;\;z=\frac{1}{\sinh(\rho)}e^{-\tau}>0.
 \label{}
\end{equation*}
In this system of coordinates the geometrical framework becomes
\begin{equation*}
\mathcal{O}=\RR_{\tau}\times\RR^3_{\mathbf x}\times]0,\infty[_{\rho},\;\; ds^2_{AdS}=\sinh^2(\rho)\left(d\tau^2-e^{2\tau}d\mathbf{x}^2\right)-d\rho^2,
 \label{}
\end{equation*}
\begin{equation*}
{\mathcal M}=\RR_{\tau}\times\RR^3_{\mathbf x}\times]0,\rho_0[_{\rho},\;\;\rho_0:=\log\left(\frac{1+\sqrt{1-\alpha^2}}{-\alpha}\right),
 \label{}
\end{equation*}
\begin{equation*}
\mathcal{B}=\RR_{\tau}\times\RR^3_{\mathbf x}\times\left\{\rho=\rho_0\right\},\;\;
ds^2_{\mathcal{B}}=\left(\frac{1}{\alpha^2}-1\right)\left(d\tau^2-e^{2\tau}d\mathbf{x}^2\right),
 \label{}
\end{equation*}
and the dynamics has the form :
\begin{equation}
\left[\frac{\partial^2}{\partial\tau^2}+3\frac{\partial}{\partial\tau}-e^{-2\tau}\Delta_{\mathbf x}-\frac{1}{\sinh^2(\rho)}\frac{\partial}{\partial\rho}\left(\sinh^4(\rho)\frac{\partial}{\partial\rho}\right)+M^2\sinh^2(\rho)\right]u=0,\;\;\tau\in\RR,\;\;{\mathbf x}\in\RR^3,\;\;0<\rho<\rho_0,
 \label{eqq}
\end{equation}
\begin{equation}
\frac{\partial u}{\partial\rho}+c u=0\;(or\; u=0\;if\;c=\infty),\;\;\tau\in\RR,\;\;{\mathbf x}\in\RR^3,\;\;\rho=\rho_0.
 \label{condlim}
\end{equation}
These coordinates are very fitted to our purpose because on the one hand, they describe exactly the manifold $\mathcal M$ in a simple form, and on the other hand we recognize in the $(\tau,\mathbf{x})$-part of (\ref{eqq}), the wave equation on the De Sitter brane $\left(\mathcal{B}, ds^2_{\mathcal B}\right)$ :
\begin{equation}
\left[\frac{\partial^2}{\partial\tau^2}+3\frac{\partial}{\partial\tau}-e^{-2\tau}\Delta_{\mathbf x}\right]\varphi=0,\;\;\tau\in\RR,\;\;{\mathbf x}\in\RR^3.
 \label{eqds}
\end{equation}
The mixed problem associated with (\ref{eqq}) and (\ref{condlim}) is solved in the next section. An important result proved in this work is the existence of the so-called Kaluza-Klein Tower, that is the representation of the finite energy solutions of  (\ref{eqq}) and (\ref{condlim}), by a superposition of massive Klein-Gordon fields propagating on the De Sitter brane. Such a result had been proved for the Minkowski brane in \cite{RS} and \cite{braneg}. In short, we show that
$$
u(\tau,\mathbf{x},\rho)=\sum_ju_{\lambda_j}(\tau,\mathbf{x})w(\rho;\lambda_j)+\int_{\frac{3}{2}}^{\infty}u_{m^2}(\tau,\mathbf{x})w(\rho;m^2)dm
$$
where $\lambda_j$ and $m^2$ describe respectively the finite point spectrum and the absolutely continuous spectrum of the Sturm-Liouville operator $L_c:=\sinh^{-2}\rho\frac{d}{d\rho}\left(\sinh^4\rho\frac{d}{d\rho}\right)+M^2\sinh^2\rho$ on $]0,\rho_0[$ completed by the boundary condition $w'(\rho_0)+cw(\rho_0)=0$, furthermore $L_cw(\rho;\kappa)=\kappa w(\rho;\kappa)$, and $u_{\kappa}(\tau,\mathbf{x})$ is a finite energy  solution of the massive Klein-Gordon equation on the De Sitter brane,
\begin{equation}
\left[\frac{\partial^2}{\partial\tau^2}+3\frac{\partial}{\partial\tau}-e^{-2\tau}\Delta_{\mathbf x}+\kappa\right]u_{\kappa}=0.
 \label{eqsted}
\end{equation}
In fact the brane is half of De Sitter space-time, it is just the Steady State Universe of Bondi, Gold and  Hoyle and we investigate the asymptotics of the solutions of (\ref{eqsted}) in the third part. A detailled spectral analysis of the operator $L_c$ is performed in the next section and allows to establish the existence of the Kaluza-Klein tower. We apply these results to the gravitational fluctuations, {\it i.e.} when $M=0$, $c=0$, in the last part where we prove that the leading term of a gravitational fluctuation near the brane is a massless graviton. The physical meaning of this property is the linear stability of the De Sitter brane. Finally we investigate the gravitational waves beyond the Cauchy horizon, and we show that the energy can blow up when the brane hits the conformal boundary of the Anti-de Sitter bulk.

We end this introduction by some bibliographical indications. Of the physical point of view, the brane cosmology is presented in the nice monography by Mannheim \cite{mannheim}, and the works of Langlois {\it et alii} \cite{langlois1}, \cite{langlois2} are fundamental references. Here are some mathematical references. Among the papers devoted to the waves equations in the  Anti-de Sitter universe, we can mention \cite{Dirac-ADS}, \cite{braneg}, \cite{supersing}, and for the general context of the asymptotically Anti-de Sitter manifolds, Holzegel \cite{holzegel}, Vasy \cite{vasy-ads} and Warnick \cite{warnick}. The wave propagation in the De Sitter like spaces has been extensively studied, in particular by Baskin \cite{baskin}, Galstian and Yagdjian \cite{galstian}, \cite{yagdjian}, Vasy \cite{vasy-ads}. The Minkowski branes are been investigated in \cite{RS}, \cite{braneg}.

\section{Propagator for the Mixed Problem}
In this section we look for a solution of (\ref{eqq}) and (\ref{condlim}) satisfying the initial condition
\begin{equation}
u(\tau_*)=u_0,\;\;\partial_{\tau}u(\tau_*)=u_1,
 \label{condinit}
\end{equation}
where the initial time $\tau_*$ is given in $\RR$ and the initial data $u_j$ belong to some suitable spaces.
To define the functional framework, we note that the smooth solutions satisfy
\begin{equation}
\begin{split}
\frac{\partial}{\partial\tau}\left(\sinh^2(\rho)\left\vert\frac{\partial u}{\partial\tau}\right\vert^2+e^{-2\tau}\sinh^2(\rho)\left\vert\nabla_{\mathbf x}u\right\vert^2+ \right. & \left.\sinh^4(\rho) \left\vert\frac{\partial u}{\partial\rho}\right\vert^2 +M^2\sinh^4(\rho)\mid u\mid^2 \right)\\
&-2\nabla_{\mathbf x}\cdot\left(e^{-2\tau}\sinh^2(\rho)\frac{\partial u}{\partial\tau}\nabla_{\mathbf x}u\right)-2\frac{\partial}{\partial\rho}\left(\sinh^4(\rho)\frac{\partial u}{\partial\tau}\frac{\partial u}{\partial\rho}\right)\\
&=-6\sinh^2(\rho)\left\vert\frac{\partial u}{\partial\tau}\right\vert^2-2e^{-2\tau}\sinh^2(\rho)\left\vert\nabla_{\mathbf x}u\right\vert^2,
\end{split}
\label{pointing}
\end{equation}
hence we consider the waves for which the following energy is well defined
\begin{equation}
\begin{split}
\mathcal{E}(u,\tau):=&\int_{\RR^3}\int_0^{\rho_0}
\sinh^2(\rho)\left\vert\frac{\partial u}{\partial\tau}\right\vert^2+e^{-2\tau}\sinh^2(\rho)\left\vert\nabla_{\mathbf x}u\right\vert^2+\sinh^4(\rho) \left\vert\frac{\partial u}{\partial\rho}\right\vert^2 +M^2\sinh^4(\rho)\mid u\mid^2 d\mathbf{x}d\rho\\
&+c\sinh^4(\rho_0)\int_{\RR^3}\mid u(\tau,\mathbf{x},\rho_0)\mid^2d\mathbf{x},
\end{split}
 \label{ener}
\end{equation}
where the last integral is missing when $c=\infty$.
Therefore it is natural to introduce the Hilbert spaces
\begin{equation}
X^0:=L^2\left(\RR^3_{\mathbf x}\times]0,\rho_0[_{\rho}, \sinh^2(\rho) d\rho d\mathbf{x}\right),\;\;\Vert u\Vert_{X^0}^2:=
\int_{\RR^3}\int_0^{\rho_0}
\sinh^2(\rho)\vert u(\mathbf{x},\rho)\vert^2 d\mathbf{x}d\rho,
 \label{xo}
\end{equation}
\begin{equation}
X^1:=\left\{u\in X^0;\;\;\nabla_{\mathbf x}u,\; \sinh(\rho)\partial_{\rho}u\in X^0\right\},\;\;\Vert u\Vert_{X^1}^2:=\Vert u\Vert_{X^0}^2+\Vert \nabla_{\mathbf x}u\Vert_{X^0}^2+\Vert  \sinh(\rho)\partial_{\rho} u\Vert_{X^0}^2
,
 \label{xun}
\end{equation}
that allow to define the previous energy since $X^1\subset C^0\left(]0,\rho_0]_{\rho};H^{\frac{1}{2}}\left(\RR^3_{\mathbf x}\right)\right)$, and the space
\begin{equation}
X^2:=\left\{u\in X^1;\;\Delta_{\mathbf x}u,\;\frac{1}{\sinh^2(\rho)}\frac{\partial}{\partial\rho}\left(\sinh^4(\rho)\frac{\partial u}{\partial\rho}\right)\in X^0\right\},
 \label{xdeu}
\end{equation}
endowed with its natural norm. For $u\in X^2$, the boundary condition (\ref{condlim}) makes sense since $X^2\subset C^1\left(]0,\rho_0]_{\rho};H^{\frac{1}{2}}\left(\RR^3_{\mathbf x}\right)\right)$ and we introduce the Hilbert subspaces :
\begin{equation*}
c\in\RR,\;\;X^2_{c}:=\left\{u\in X^2;\;\;\partial_{\rho}u(\rho_0)+c u(\rho_0)=0\right\},\;\;X^2_{\infty}:=\left\{u\in X^2;\;u(\rho_0)=0\right\}.
 \label{}
\end{equation*}
When $c=\infty$, the homogeneous Dirichlet problem is investigated in the closed subspace
\begin{equation*}
X^1_0:=\left\{ u\in X^1;\;\;u(\rho_0,.)=0\right\}.
 \label{}
\end{equation*}

The solutions  of (\ref{eqq}) in $ C^2\left(\RR_{\tau};X^0\right)\cap C^1\left(\RR_{\tau};X^1\right)\cap C^0\left(\RR_{\tau};X^2_{c}\right)$ are called {\it strong solutions} of (\ref{eqq}) and (\ref{condlim}). To give a meaning to the boundary condition for the finite energy solutions in $ C^1\left(\RR_{\tau};X^0\right)\cap C^0\left(\RR_{\tau};X^1\right)$ we have to introduce a suitable space of distributions on $]0,\rho_0[$.

 The partial differential equation (\ref{eqq}) can we written as
\begin{equation}
\partial_{\tau}^2u+3\partial_{\tau}u+A_{\tau}u=0,\;\;A_{\tau}:=-e^{-2\tau}\Delta_{\mathbf x}+L,
 \label{zeqq}
\end{equation}
where $L$ is the differential operator
\begin{equation*}
L:=-\frac{1}{\sinh^2(\rho)}\frac{\partial}{\partial\rho}\left(\sinh^4(\rho)\frac{\partial}{\partial\rho}\right)+M^2\sinh^2(\rho),\;\;\rho\in J:=]0,\rho_0[.
 \label{}
\end{equation*}
First we investigate $L$ considered as a Sturm-Liouville operator 
on the Hilbert space
\begin{equation*}
H:=L^2(J,\sinh^2(\rho) d\rho).
 \label{}
\end{equation*}
We apply classical results on the spectral analysis of the differential operators (the fundamental references are \cite{naimark} and \cite{weidmann}, see also \cite{zettl}).
First, we introduce the maximal/minimal domains
\begin{equation*}
D_{max}:=\left\{u\in H; Lu\in H\right\},\;\;D'_{min}:=\left\{u\in D_{max}; u\; has\;compact\;support\;in\; J\right\}.
 \label{}
\end{equation*}
For all $\epsilon>0$, and $u\in D_{max}$, the restriction of $u$ to $ ]\epsilon,\rho_0[$ belongs to $H^2(]\epsilon,\rho_0[)$ hence $u(\rho_0)$ and $u'(\rho_0)$ are well defined and for $c\in\RR$ and $c=\infty$ we can introduce
\begin{equation*}
D_{c}:=\left\{u\in D_{max};\;u'(\rho_0)+cu(\rho_0)=0\right\},\;\;D_{\infty}:=\left\{u\in D_{max};\;u(\rho_0)=0\right\}.
 \label{}
\end{equation*}
We define $L_{c}$ (respectively $L'_{min}$, $L_{max}$)  as the operator $L$ endowed with the domain $D_{c}$ (respectively $D'_{min}$, $D_{max}$). We denote $L_{min}$ the closure of $L'_{min}$. We know that
$D'_{min}$ is dense in $H$, and
\begin{equation*}
L'_{min}\subset L_{min}\subset L^*_{min}=L'^*_{min}=L_{max},\;\;L^*_{max}=L_{min}.
 \label{}
\end{equation*}

It is obvious that $\rho_0$ is a regular point since $\sinh(\rho)$ is a continuous function and $\sinh(\rho_0)>0$. Furthermore $0$ is in the point limit case and this result follows from Theorem 6.3 of  \cite{weidmann} since for $0<\rho_1<\rho_0$ we have
$$
\int_{\rho}^{\rho_1}\frac{1}{\sinh^4(r)}dr\sim_{\rho\rightarrow 0}\rho^{-3}\notin L^2(]0,\rho_1[,\sinh^2(\rho)d\rho).
$$
Moreover we know that for all $u,v\in D_{max}$ the following limit exists at zero:
\begin{equation*}
[u,v]_0:=\lim_{\rho\rightarrow 0}u'(\rho)v(\rho)-u(\rho)v'(\rho).
 \label{}
\end{equation*}
Now the theorems 3.12 and 5.7 of \cite{weidmann} assure that the domain $D_{min}$ of $L_{min}$ is characterized by
\begin{equation*}
D_{min}=\left\{u\in D_{max};\;\;u(\rho_0)=u'(\rho_0)=0,\;\;[u,v]_0=0\;for\;every\;v\in D_{max}\right\},
 \label{}
\end{equation*}
and its deficiency indices $\gamma_{\pm}\left(L_{min}\right):= dim\,Ran\left(z-L_{min}\right)^{}=dim\,Ker\left(\bar{z}-L_{min}\right)$, $z\in\CC$, $\pm \Im z>0$, are equal to $1$. Finally we deduce from theorem 5.8 of \cite{weidmann} the set of all the self-adjoint extensions of $L_{min}$ :


\begin{Lemma}
 \label{}
For all $c\in\RR\cup\{\infty\}$, the operator $L_{c}$ is self-adjoint on $H$. Conversely, any self-adjoint extension of $L_{min}$ has the form $L_{c}$ for some  $c\in\RR\cup\{\infty\}$.
\end{Lemma}


To define the weak solutions, we denote ${\mathcal D}'\left(\RR_{\tau}\times\RR^3_{\mathbf x};D_{c}\right)$ the space of the vector distributions on $\RR_{\tau}\times\RR^3_{\mathbf x}$, that are $D_{c}$-valued, and we call {\it finite energy solutions} of (\ref{eqq}) and (\ref{condlim}), the distributions that are  solutions of (\ref{eqq}) and belong to $ C^1\left(\RR_{\tau};X^0\right)\cap C^0\left(\RR_{\tau};X^1\right)\cap{\mathcal D}'\left(\RR_{\tau}\times\RR^3_{\mathbf x};D_{c}\right)$.


The main result of this part is the following :
\begin{Theorem}
 \label{theopropag}
 Given $M\geq 0$, $\alpha\in]-1,0[$, $c\in\RR$, $u_0\in X^1$ and $u_1\in X^0$, there exists a unique finite energy solution $u$ of (\ref{eqq}) and (\ref{condinit}). The energy (\ref{}) is decreasing and there exists $f\in C^0(\RR^+;\RR^+)$ such that for all $\tau\in\RR$ we have
\begin{equation}
\Vert u(\tau)\Vert_{X^1}+\Vert \partial_{\tau}u(\tau)\Vert_{X^0}\leq f(\mid\tau-\tau_*\mid)\left(\Vert u_0\Vert_{X^1}+\Vert u_1\Vert_{X^0}\right),
 \label{estimreg}
\end{equation}
and there exists $C\in C^0\left( \RR_{\tau_*}\times C^{\infty}_0(\RR_{\tau}\times\RR^3_{\mathbf{x}});\RR^+\right)$ such that for all
 $\Theta\in C^{\infty}_0(\RR_{\tau}\times\RR^3_{\mathbf{x}})$,
\begin{equation}
\Vert\int\Theta(\tau,\mathbf{x})u(\tau,\mathbf{x},.)d\tau d\mathbf{x}\Vert_{D_{c}}\leq C(\tau_*,\Theta)\left(\Vert u_0\Vert_{X^1}+\Vert u_1\Vert_{X^0}\right).
 \label{estimregde}
\end{equation}
Here, $f$ and $C$ are independent of $u_0$, $u_1$, and only depends on $M$, $\alpha$ and $c$.\\

If $u_0\in X^2_{c}$ and $u_1\in X^1$, then $u$ is a strong solution, {\it i.e.} $u\in C^2\left(\RR_{\tau};X^0\right)\cap C^1\left(\RR_{\tau};X^1\right)\cap C^0\left(\RR_{\tau};X^2_{c}\right)$.\\

All the previous results hold when $c=\infty$ by replacing $X^1$ by $X^1_0$.
\end{Theorem}


{\it Proof of Theorem \ref{theopropag}.} The demonstration is based on the famous result of T. Kato on the propagator of the time-dependent hyperbolic evolution equations (see {\it e.g.} \cite{tanabe}).
It is convenient to construct the functional framework to solve the resolvent equation
\begin{equation}
L_{c}u+iu=f.
 \label{resolveq}
\end{equation}
If $c\in\RR$, we introduce the space
\begin{equation}
H_1:=\left\{u\in H;\;\sinh(\rho) u'\in H\right\},\;\;\Vert u\Vert_{H_1}^2:=\Vert u\Vert_{H}^2+\Vert \sinh(\rho) u'\Vert_{H}^2,
 \label{hun}
\end{equation}
and the following sesquilinear form
\begin{equation*}
a_{c}(u,v):=\int_0^{\rho_0}\sinh^4(\rho) u'\overline{v'}+M^2\sinh^4(\rho) u\overline{v}+i\sinh^2(\rho) u\overline{v}\; d\rho+c\sinh^4(\rho_0)u(\rho_0)\overline{v(\rho_0)}.
 \label{}
\end{equation*}
It is obvious that $a_{c}$ is continuous on $H_1$ and 
\begin{equation}
\Im a_{c}(u,u)=\Vert u\Vert^2_H.
 \label{coerh}
\end{equation}
Now we show that for any $\epsilon>0$ we can choose $C_{\epsilon}>0$ such that
\begin{equation}
\mid u(\rho_0)\mid^2\leq C_{\epsilon} \Vert u\Vert^2_H+\epsilon \int_0^{\rho_0}\sinh^4(\rho) \mid u'(\rho)\mid^2d\rho.
 \label{coerh'}
\end{equation}
We write for $\rho\in]\frac{\rho_0}{2},\rho_0[$, $\epsilon>0$,
$$
\mid u(\rho_0)\mid^2=\mid u(\rho)\mid^2+2\Re \int_{\rho}^{\rho_0}u(r)\overline{u'(r)}dr
\leq \mid u(\rho)\mid^2+\epsilon\int_{\frac{\rho_0}{2}}^{\rho_0}\mid u'(r)\mid^2dr+\frac{1}{\epsilon}\int_{\frac{\rho_0}{2}}^{\rho_0}\mid u(r)\mid^2dr
$$
and we integrate with respect to $\rho$ between $\frac{\rho_0}{2}$ and $\rho_0$ :
$$
\mid u(\rho_0)\mid^2
\leq \frac{2}{\rho_0}\int_{\frac{\rho_0}{2}}^{\rho_0}\mid u(\rho)\mid^2d\rho+\epsilon\int_{\frac{\rho_0}{2}}^{\rho_0}\mid u'(r)\mid^2dr+\frac{1}{\epsilon}\int_{\frac{\rho_0}{2}}^{\rho_0}\mid u(r)\mid^2dr,
$$
and we finally get :
$$
\mid u(\rho_0)\mid^2
\leq \left( \frac{2}{\rho_0}+\frac{1}{\epsilon}\right)\frac{1}{\sinh^2\left(\frac{\rho_0}{2}\right)}\Vert u\Vert^2_H+
\frac{\epsilon}{\sinh^4\left(\frac{\rho_0}{2}\right)} \int_0^{\rho_0}\sinh^4(\rho) \mid u'(\rho)\mid^2d\rho.
$$

Using (\ref{coerh'}) we deduce that $a_{c}$ is coercive on $H_1$ and so the Lax-Milgram theorem assures that for any $f\in H$ there exists a unique $u_f\in H_1$ such that $a_{c}(u_f,v)=<f,v>_H$ for all $v\in H_1$. Taking $v\in C^{\infty}_0(]0,\rho_0[)$, and then $v\in C^{\infty}_0(]0,\rho_0])$, it is easy to check that $u_f\in D_{c}$ and $u_f$ is solution of (\ref{resolveq}). When $c=\infty$, we use the sesquilinear form $a_0$ on the Hilbert space
\begin{equation*}
\dot{H}_1:=\left\{u\in H_1;\;\;u(\rho_0)=0\right\}.
 \label{}
\end{equation*}
As a consequence of this construction, we get that if $u\in D_{c}$ we have $u=u_f$ for $f=Lu+iu$, and we deduce that $D_{c}\subset H_1$. Since $D_{max}$ is obviously the union of all the $D_{c}$ for $c\in\RR\cup\{\infty\}$ we conclude that
\begin{equation*}
D_{max}=D_{max}\cap H_1,
 \label{}
\end{equation*}
and since $D_{max}$ and $D_{max}\cap H_1$ are two Hilbert spaces for their natural norms, the Banach theorem assures that these norms are equivalent et there exists $C>0$ such that
\begin{equation}
\forall u\in D_{max},\;\;\Vert u\Vert_{H_1}\leq C\left(\Vert u\Vert_H+\Vert Lu\Vert_H\right).
 \label{interpol}
\end{equation}
Furthermore we obtain a characterization of $X^2$ :
\begin{equation}
X^2=\left\{u\in X^0;\;\Delta_{\mathbf x}u,\;\frac{1}{\sinh^2(\rho)}\frac{\partial}{\partial\rho}\left(\sinh^4(\rho)\frac{\partial u}{\partial\rho}\right)\in X^0\right\}.
 \label{xdeuzo}
\end{equation}
Also we deduce from this variational approach that the Green formula is valid for  all $u\in D_{max}$, $v\in H_1$ :
\begin{equation}
\left<Lu,v\right>_H=\int_0^{\rho_0}\sinh^4(\rho) u'\overline{v'}+M^2\sinh^4(\rho) u\overline{v} d\rho-\sinh^4(\rho_0)\partial_{\rho}u(\rho_0)\overline{v(\rho_0)}.
 \label{green}
\end{equation}
In particular for $u\in D_{c}$ we have :
\begin{equation}
\left<L_{\gamma}u,u\right>_H=\int_0^{\rho_0}\left(\mid \partial_{\rho}u\mid^2+M^2\mid u \mid^2\right)\sinh^4(\rho) d\rho+c\sinh^4(\rho_0)\mid u(\rho_0)\mid^2,\;if\;c\in\RR,
 \label{greengamma}
\end{equation}
\begin{equation}
\left<L_{-\infty}u,u\right>_H=\int_0^{\rho_0}\left(\mid \partial_{\rho}u\mid^2+M^2\mid u \mid^2\right)\sinh^4(\rho) d\rho.
 \label{greendiri}
\end{equation}
These equalities could also be obtained by an integration by parts, with some sharp asymptotics estimates of $u(\rho)$ and $u'(\rho)$ as $\rho\rightarrow 0$, for $u\in D_{max}$ (see (\ref{ureau}) below). Another consequence of these formulas and (\ref{coerh}), (\ref{coerh'}) is that 
there exists $A,C>0$ such that for any $u\in D_c$ we have :
\begin{equation}
C^{-1}\Vert u\Vert^2_{H_1}\leq \left<L_cu,u\right>_H+A\Vert u\Vert^2_H\leq C \Vert u\Vert^2_{H_1}
 \label{ekivh}
\end{equation}
Finally we also note that $L_{c}$ is bounded from below and for all $M\geq 0$, there exists $c_M\in[0,\infty[$ such that
\begin{equation*}
\forall c\in[c_M,\infty],\;\;0\leq L_{c}.
 \label{}
\end{equation*}

We now return to the hyperbolic equation (\ref{zeqq}) that we express as
\begin{equation}
\frac{\partial}{\partial\tau}
U(\tau)=[\AA(\tau) +\BB]U(\tau),\;\;U:= \left(
\begin{array}{c}
u\\
\partial_{\tau}u
\end{array}
\right),
\;\;
\AA(\tau):=
\left(
\begin{array}{cc}
0&1\\
-A_{\tau}-\Gamma&0
\end{array}
\right),
\;\;
\BB:=\left(
\begin{array}{cc}
0&0\\
\Gamma&-3
\end{array}
\right),
 \label{systeq}
\end{equation}
where  using (\ref{coerh'}) we have taken $\Gamma>1$ large enough to that $1\leq L_c+\Gamma$ and
$$
\mid c\mid\sinh^4(\rho_0)\int_{\RR^3}\mid u(\mathbf{x},\rho_0)\mid^2d\mathbf{x}\leq \int_{\RR^3}\int_0^{\rho_0} \left[\frac{1}{2}\sinh^2(\rho) \left\vert\frac{\partial u}{\partial\rho}\right\vert^2+(\Gamma-1) \mid u\mid^2\right] \sinh^2(\rho) d\mathbf{x}d\rho.
$$
Now given $\tau\in\RR$, we consider $\AA(\tau)$ as an operator on $X^1\times X^0$ endowed with the equivalent norm
\begin{equation*}
\begin{split}
\left\Vert (u,v)\right\Vert_{\tau}^2:=&\int_{\RR^3}\int_0^{\rho_0}
\left[\left\vert v\right\vert^2+e^{-2\tau}\left\vert\nabla_{\mathbf x}u\right\vert^2+\sinh^2(\rho) \left\vert\frac{\partial u}{\partial\rho}\right\vert^2 +\left(M^2\sinh^2 (\rho)+\Gamma\right)\mid u\mid^2\right] \sinh^2(\rho) d\mathbf{x}d\rho\\
&+c\sinh^4(\rho_0)\int_{\RR^3}\mid u(\mathbf{x},\rho_0)\mid^2d\mathbf{x}.
\end{split}
 \label{}
\end{equation*}
 and we define the domain of $\AA(\tau)$ as $Dom(\AA(\tau))=X^2_c\times X^1$. Thanks to (\ref{green}), we easily check that for all $U,V\in Dom(\AA(\tau))$, we have
$$
\left<i\AA(\tau)U;V\right>_{\tau}=\left<U;i\AA(\tau)V\right>_{\tau}.
$$
Now given $\epsilon=\pm 1$, $V=(f,g)\in X^1\times X^0$, we show that the equation
$$
\AA(\tau)U+\epsilon U=V
$$
has a unique solution $U=(u_{\epsilon},v_{\epsilon})\in Dom(\AA(\tau))$. $U$ is solution iff $v_{\epsilon}=f-\epsilon u_{\epsilon}$ and $u_{\epsilon}\in X^2_c$ is solution of
\begin{equation}
-e^{-2\tau}\Delta_{\mathbf x}u_{\epsilon}+Lu_{\epsilon}+(\Gamma+1)u_{\epsilon}=\epsilon f-g=:h_{\epsilon}\in X^0,
 \label{eqfg}
\end{equation}
and this last equation is easily solved by the Lax-Milgram lemma applied to the variational problem in $X^1$:
\begin{equation}
\begin{split}
\forall u'\in X^1,\;\;\int_{\RR^3}\int_0^{\rho_0}
&\left[e^{-2\tau}\nabla_{\mathbf x}u_{\epsilon}\overline{\nabla_{\mathbf x}u'}+\sinh^2(\rho) \partial_{\rho}u_{\epsilon}\overline{\partial_{\rho}u'} +\left(M^2\sinh^2 (\rho)+\Gamma+1\right)u_{\epsilon}\overline{u'}\right] \sinh^2(\rho) d\mathbf{x}d\rho\\
&+c\sinh^4(\rho_0)\int_{\RR^3}u_{\epsilon}(\mathbf{x},\rho_0)\overline{u'(\mathbf{x},\rho_0)}d\mathbf{x}=\int_{\RR^3}\int_0^{\rho_0}h_{\epsilon}\overline{u'}\sinh^2(\rho) d\mathbf{x}d\rho.
\end{split}
 \label{pbvarx}
\end{equation}
This problem has a unique solution $u_{\epsilon}\in X^1$, and taking $u'\in C^{\infty}_0\left(\RR^3_{\mathbf x}\times]0,\rho_0[\right)$ we get that $u_{\epsilon}$ is solution of (\ref{eqfg}). We deduce that for any $\rho_1\in]0,\rho_0[$, we have $\partial_{\rho}u_{\epsilon}\in L^2\left(]\rho_1,\rho_0[_{\rho}; L^2(\RR^3_{\mathbf x})\right)$ and $\partial^2_{\rho}u_{\epsilon}\in L^2\left(]\rho_1,\rho_0[_{\rho}; H^{-1}(\RR^3_{\mathbf x})\right)$. The theorem of the traces (\cite{lions}, page 23) assures that $\partial_{\rho}u_{\epsilon}\in C^0\left([\rho_1,\rho_0]_{\rho}; H^{-\frac{1}{2}}(\RR^3_{\mathbf x})\right)$.
Taking  $u'\in C^{\infty}_0\left(\RR^3_{\mathbf x}\times]0,\rho_0]\right)$ in (\ref{pbvarx}) and using equation (\ref{eqfg}), we conclude that
$$
\left<\partial_{\rho}u_{\epsilon}(\rho_0,.)+c u_{\epsilon}(\rho_0,.);u'(\rho_0,.)\right>_{H^{-\frac{1}{2}}(\RR^3_{\mathbf x}),H^{\frac{1}{2}}(\RR^3_{\mathbf x})}=0
$$
and therefore $u_{\epsilon}$ satisfies the boundary condition $\partial_{\rho}u_{\epsilon}(\rho_0,.)=\gamma u_{\epsilon}(\rho_0,.)$. Then the theorem of elliptic regularity for the Neumann problem assures that $u_{\epsilon}\in H^2(]\rho_1,\rho_0[\times \RR^3)$ but it remains to prove that it belongs to $X^2$. We denote $\widehat{u_{\epsilon}}(\rho,\pmb{\xi})$ the partial Fourier transform with respect to $\mathbf x$ of $u_{\epsilon}$. We have :
$$
\widehat{u_{\epsilon}},\;e^{-2\tau}\mid\pmb{\xi}\mid^2 \widehat{u_{\epsilon}}+L \widehat{u_{\epsilon}}+(\Gamma+1) \widehat{u_{\epsilon}}\in L^2\left(]0,\rho_0[_{\rho}\times\RR^3_{\pmb\xi}, \sinh^2(\rho)d\rho d\pmb{\xi}\right),\;\;\partial_{\rho}\widehat{u_{\epsilon}}(\rho_0,.)=\gamma \widehat{u_{\epsilon}}(\rho_0,.),
$$
hence for almost all $\pmb{\xi}\in\RR^3$, the map $\rho\mapsto \widehat{u_{\epsilon}}(\rho,\pmb{\xi})$ belongs to $D_{c}$. The Parseval equality and the theorem of Fubini allow to write
$$
e^{-4\tau}\Vert\Delta_{\mathbf x}u_{\epsilon}\Vert^2_{X^0}+\Vert (L+\Gamma+1)u_{\epsilon}\Vert^2_{X^0}+2e^{-2\tau}\Re\int_{\RR^3}\mid\pmb\xi\mid^2\left<(L_{c}+\Gamma+1) \widehat{u_{\epsilon}}(.,\pmb{\xi}),\widehat{u_{\epsilon}}(.,\pmb{\xi})\right>_Hd\pmb{\xi}=\Vert h_{\epsilon}\Vert^2_{X^0}<\infty.
$$
Since $L_{c}+\Gamma+1$ is a positive operator, we deduce that $\Delta_{\mathbf x}u_{\epsilon}$ and $Lu_{\epsilon}$ belong to $X^0$ and therefore $u_{\epsilon}\in X^2_{c}$.

We conclude that $\left(i\AA(\tau), Dom\left(i\AA(\tau)\right)=X^2_{c}\times X^1\right)$ is a densely defined selfadjoint operator in $\left(X^1\times X^0, \Vert.\Vert_{\tau}\right)$, hence $\AA(\tau)$ and $-\AA(\tau)$ generate $C^0$ unitary groups on $\left(X^1\times X^0, \Vert.\Vert_{\tau}\right)$ leaving invariant $X^2_{c}\times X^1$. Since we obviously have :
$$
\tau_1\leq\tau_2\Rightarrow \Vert.\Vert_{\tau_2}\leq \Vert.\Vert_{\tau_1}\leq e^{\tau_2-\tau_1}\Vert.\Vert_{\tau_2},
$$
we deduce that for any $U\in X^1\times X^0$, 
$T>0$, $t_j\geq 0$ and $\tau_j$ with $0=\tau_0\leq \tau_1\leq \tau_2\leq ...\leq \tau_n\leq T$, we have
$$
\Vert e^{t_k\AA(\tau_*+\tau_k)}U\Vert_{\tau_*\tau_k}\leq e^{\tau_k-\tau_{k-1}}\Vert U\Vert_{\tau_*\tau_{k-1}},
$$
hence by iteration
$$
\left\Vert e^{t_n\AA(\tau_*+\tau_n)}e^{t_{n-1}\AA(\tau_*+\tau_{n-1})}...e^{t_1\AA(\tau_*+\tau_1)}U\right\Vert_{\tau_*}\leq e^{2T}\Vert U\Vert_{\tau_*},
$$
$$
\left\Vert e^{-t_n\AA(\tau_*-\tau_n)}e^{-t_{n-1}\AA(\tau_*-\tau_{n-1})}...e^{-t_1\AA(\tau_*-\tau_1)}U\right\Vert_{\tau_*}\leq e^{2T}\Vert U\Vert_{\tau_*}.
$$
We conclude that $\AA(\tau_*+\tau)$ and $-\AA(\tau_*-\tau)$ for $\tau\in[0,T]$ are stable families of infinitesimal generators with stability constants $e^{2T}$ and $0$.
Moreover, since  $\BB$ is a bounded operator on $\left(X^1\times X^0, \Vert.\Vert_{c}\right)$ with a norm independent of $\tau$, and on $X^2_{c}\times X^1$, Proposition 7.4 of \cite{tanabe} assures that for $\tau\in[0,T]$, $\pm(\AA(\tau_*\pm\tau)+\BB)$ are also stable families of infinitesimal generators in $X^1\times X^0 $ with the same domain $X^2_{c}\times X^1$, and stability constants $e^{2T}$ and $\beta(T):=\Vert\BB\Vert_{\mathcal{L}(X^1\times X^0)} e^{2T}$. Finally, given $U_0\in X^2_{c}\times X^1$, the maps $\tau\mapsto \pm(\AA(\tau_*\pm\tau)+\BB)U_0$ are obviously strongly differentiable in $X^1\times X^0$. Therefore the hypotheses of the Kato theorem are fulfilled (see {\it e.g.} Theorem 7.4 in \cite{tanabe}) and there exists a unique solution $U\in C^0\left([\tau_*-T,\tau_*+T];X^2_{c}\times X^1\right)\cap C^1\left[\tau_*-T,\tau_*+T];X^1\times X^0\right)$ solution of (\ref{systeq}) satisfying $U(\tau_*)=U_0$ and 
$$
\left\Vert U(\tau)\right\Vert_{X^1\times X^0}\leq e^{2T}e^{\beta(T)\mid \tau-\tau_*\mid}\left\Vert U_0\right\Vert_{X^1\times X^0},\;\;\tau_*-T\leq \tau\leq\tau_*+T,\;\;\forall T>0.
$$
We deduce that the strong solution exists and is unique, and also (\ref{estimreg}) holds with $f(\sigma)=e^{2\sigma+\beta(\sigma)\sigma}$. We note that $\beta$ only depends on $M$ and $\Gamma$, where $\Gamma$ is choosen just depending on $M$ and $c$. Furthermore, since $C^0\left(\RR_{\tau};X^2_{\gamma}\right)\subset L^2_{loc}\left(\RR_{\tau}\times\RR^3_{\mathbf x};D_{c}\right)$, given $\Theta\in C^{\infty}_0(\RR_{\tau}\times\RR^3_{\mathbf{x}})$, a strong solution $u$ satisfies
$$
L\left(\int\Theta(\tau,\mathbf{x})u(\tau,\mathbf{x},.)d\tau d{\mathbf x}\right)=
\int u(\tau,\mathbf{x},.)\left[-\partial_{\tau}^2\Theta+3\partial_{\tau}\Theta+e^{-2\tau}\Delta\Theta\right](\tau,\mathbf{x}) d\tau d\mathbf{x}
$$
and taking the $H$-norm and using the Cauchy-Schwarz inequality and (\ref{estimreg}) we obtain (\ref{estimregde}) with
$$
C(\tau_*,\Theta):=\int f(\tau-\tau_*)\left\Vert\left[-\partial_{\tau}^2\Theta+3\partial_{\tau}\Theta+e^{-2\tau}\Delta\Theta\right](\tau,.)\right\Vert_{L^2(\RR^3_{\mathbf{x}})}d\tau.
$$
To get the decay of the energy, it is sufficient to check that its $\tau$-derivative is negative (we also could integrate (\ref{pointing}) on $[\tau_*,\tau]\times\RR^3_{\mathbf x}\times [0,\rho_0]$).
Now the existence of the weak solutions is a straight consequence of estimates  (\ref{estimreg}) and (\ref{estimregde}).
Finally we establish the uniqueness result. We already have mentioned the uniqueness of  the strong solutions. We consider a weak solution  $u\in C^1\left(\RR_{\tau};X^0\right)\cap C^0\left(\RR_{\tau};X^1\right)\cap{\mathcal D}'\left(\RR_{\tau}\times\RR^3_{\mathbf x};D_{c}\right)$ solution of (\ref{eqq}) with $u(\tau_*)=\partial_{\tau}u(\tau_*)=0$. We take $\tau_1\in \RR$ and we have to prove that $u(\tau_1)=0$. We pick functions $\theta\in C^{\infty}_0(\RR)$, $\Phi\in C^{\infty}_0(\RR^3)$ such that $0\leq \theta,\Phi$, $\int_{\RR}\theta(\tau)d\tau=\int_{\RR^3}\Phi(\mathbf{x})d\mathbf{x}=1$, and for all integer $n$ we introduce $\theta_n(\tau):=n\theta(n\tau)$, $\Phi_n(\mathbf{x}):=n^3\Phi(n\mathbf{x})$,
$$
u_n(\tau,\mathbf{x},\rho):=\int_{\RR\times\RR^3}\theta_n(\tau-\sigma)\Phi_n(\mathbf{x}-\mathbf{y})u(\sigma,\mathbf{y},\rho)d\sigma d\mathbf{y},
$$
$$
v_n(\tau,\mathbf{x},\rho):=\int_{\RR\times\RR^3}e^{2(\tau-\sigma)}\theta_n(\tau-\sigma)\Phi_n(\mathbf{x}-\mathbf{y})u(\sigma,\mathbf{y},\rho)d\sigma d\mathbf{y}.
$$
It is easy to check that $u_n,\,v_n\in C^{\infty}\left(\RR_{\tau};X^2_{c}\right)$ and $u_n$ and $v_n$ tend to $u$ in $C^1\left(\RR_{\tau};X^0\right)\cap C^0\left(\RR_{\tau};X^1\right)$ as $n\rightarrow\infty$. We consider $w_1\in X^1$ and $w\in C^2\left(\RR_{\tau};X^0\right)\cap C^1\left(\RR_{\tau};X^1\right)\cap C^0\left(\RR_{\tau};X^2_{c}\right)$ the unique strong solution of (\ref{eqq}) with $w(\tau_1)=0$, $\partial_{\tau}w(\tau_1)=w_1$. We put
$$
f(\tau):=\left<u(\tau);\partial_{\tau}w(\tau)\right>_{X^0}-\left<\partial_{\tau}u(\tau);w(\tau)\right>_{X^0},\;\;
f_n(\tau):=\left<u_n(\tau);\partial_{\tau}w(\tau)\right>_{X^0}-\left<\partial_{\tau}u_n(\tau);w(\tau)\right>_{X^0}.
$$
We have $f(\tau_*)=0$, $f(\tau_1)=\left<u(\tau_1);w_1\right>_{X^0}$, $f\in C^0(\RR)$, $f_n\in C^{\infty}(\RR)$, and $f_n\rightarrow f$ in $C^0(\RR)$ as $n\rightarrow\infty$. Now we calculate :
\begin{equation*}
\begin{split}
f'_n(\tau)=&\left<u_n(\tau);\partial_{\tau}^2w(\tau)\right>_{X^0}-\left<\partial_{\tau}^2u_n(\tau);w(\tau)\right>_{X^0}\\
=&\left<u_n(\tau);-3\partial_{\tau}w(\tau)+e^{-2\tau}\Delta_{\mathbf x}w(\tau)-L_{\gamma}w(\tau)\right>_{X^0}\\
&-\left<-3\partial_{\tau}u_n(\tau)+e^{-2\tau}\Delta_{\mathbf x}v_n(\tau)-L_{\gamma}u_n(\tau) ;w(\tau)\right>_{X^0}\\
=&-3f_n(\tau)+e^{-2\tau}\left<\nabla_{\mathbf x}u_n(\tau)-\nabla_{\mathbf x}v_n(\tau);\nabla_{\mathbf x}w(\tau)\right>_{X^0}.
\end{split}
\end{equation*}
To obtain the last equality we have used the Green formula, the Fubini theorem and the self-adjointness of $L_{c}$. Then we get
$$
f_n(\tau_1)=f_n(\tau_*)+e^{-3\tau_1}\int_{\tau_*}^{\tau_1}e^{\sigma}\left<\nabla_{\mathbf x}u_n(\sigma)-\nabla_{\mathbf x}v_n(\sigma);\nabla_{\mathbf x}w(\sigma)\right>_{X^0}d\sigma\longrightarrow 0,\;n\rightarrow\infty.
$$
We deduce that $f(\tau_1)=0$ and since $w_1$ is arbitrarily choosen we conclude that $u(\tau_1)=0$.
Finally, all the previous proofs hold for the more simple case of the Dirichlet boundary condition ($c=\infty$), by replacing $X^1$ by $X^1_0$.
\fin

We end this part by noting that it would be possible to solve the mixed problem in larger framework, and obtain very weak solutions in $C^0\left(\RR;X^0\right)$ by using the technics of Lions-Magenes (\cite{lions}, chapter 3, sections 8 and 9).


\section{Asymptotics for the Klein-Gordon equation in the Steady State Universe $dS^4_{\frac{1}{2}}$}
The Steady State Universe is half of the 3+1 dimensional De Sitter space-time,
$$
dS_{\frac{1}{2}}^4:=\RR_{\tau}\times\RR^3_{\mathbf x},\;\; g_{\mu\nu}dx^{\mu}dx^{\nu}=d\tau^2-e^{2\tau}d\mathbf{x}^2,
$$





\begin{figure}
\begin{tikzpicture}
\draw [-] (0,4) -- (6,4)  node[midway,sloped,above] {$\tau=+\infty$} ;
\draw [-] (6,4) -- (6,-2) ;
\draw [-] (6,-2) -- (0,-2) ;
\draw [-] (0,-2) -- (0,4) node[midway,sloped,above] {$\mathbf{x}=0$};
\draw (0,-2) -- (6,4) node[midway,sloped,above] {$\tau=-\infty$};
\draw (7,4.3) node {$i_0$ ($\mid\mathbf{x}\mid=\infty)$};
\draw plot[smooth] coordinates{(0,1.5) (2,2) (6,4)};
\draw plot[smooth] coordinates{(0,-2) (2,2) (2.5,4)};
\draw (1,0.5) node[rotate=63]{$\mid\mathbf{x}\mid=R$};
\draw (4,3.1) node[rotate=28]{$\tau=\tau_*$};
\draw (2,2) node {$\bullet$};
\draw (4,4) node {$\bullet$};
\draw (1.8,2.2) node {A};
\draw (4.2,4.2) node{B};
\draw [-] (2,2) -- (4,4);
\fill[color=gray!20]
(0,-2) -- (6,4)
-- (6,-2)--(0,-2)
-- cycle;
\end{tikzpicture}
\caption{The whole square is the De Sitter space-time. The white part is the Steady-State Universe. Each point is a 2-sphere. The null lines are at $45^0$.  If $A$ is the sphere of radius $R$ located at $\tau=\tau_*$, its horizon at $\tau=\infty$ is the two sphere $B$ of radius $R+e^{-\tau_*}$.}
\label{crobasteady}
\end{figure}
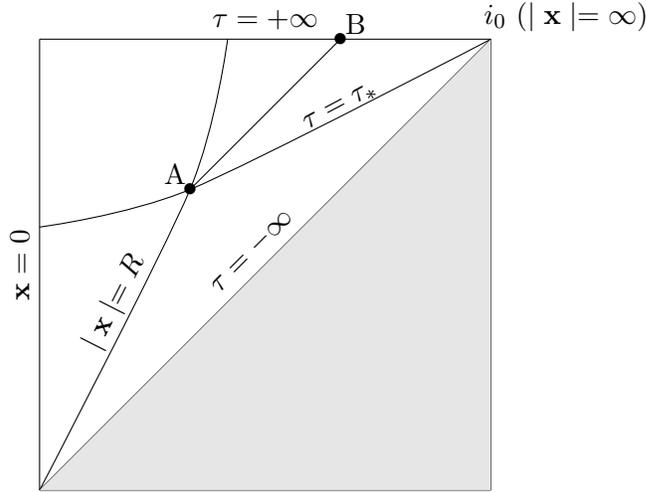


On this manifold, the Klein-Gordon equation $\frac{1}{\sqrt{\mid g\mid}}\frac{\partial}{\partial x^{\mu}}\left(\sqrt{\mid g\mid}g^{\mu \nu}\frac{\partial}{\partial x^{\nu}}\right)u+\kappa u=0$, $\kappa\in\CC$, has the form
\begin{equation}
\left[\frac{\partial^2}{\partial\tau^2}+3\frac{\partial}{\partial\tau}-e^{-2\tau}\Delta_{\mathbf x}+\kappa\right]u=0,\;\;\tau\in\RR,\;\;{\mathbf x}\in\RR^3.
 \label{kgdsu}
\end{equation}
Since we deal with the wave equation on a globally hyperbolic lorentzian $C^{\infty}$ manifold, it is well-known that the global Cauchy problem is well posed in $C^{\infty}_0\left(\RR^3_{\mathbf x}\right)$ and in $\mathcal{D}'\left(\RR^3_{\mathbf x}\right)$, and the fundamental solution has been computed in \cite{galstian} when $\kappa\geq 0$ and \cite{yagdjian} for any $\kappa$. In this part, we first prove the Cauchy problem is well posed in the scale of the usual Sobolev spaces, {\it i.e.}
given $s\in\RR$, we want to look for the solution
\begin{equation}
u\in C^0\left(\RR_{\tau};H^s\left(\RR^3_{\mathbf x}\right)\right)\cap C^1\left(\RR_{\tau};H^{s-1}\left(\RR^3_{\mathbf x}\right)\right),
 \label{regliss}
\end{equation}
of (\ref{kgdsu}), satisfying at some time $\tau_*\in\RR$,
\begin{equation}
u(\tau_*,.)=u_0(.)\in H^s\left(\RR^3_{\mathbf x}\right),\;\;\partial_{\tau}u(\tau_*,.)=u_1(.)\in H^{s-1}\left(\RR^3_{\mathbf x}\right),
 \label{initialdsu}
\end{equation}
and depending continuously of these initial data. This result is not at all surprising and there are a lot of possible strategies: Kato's theorem, transposition method {\it \`{a} la} Lions \cite{lions}, etc., but we adopt a pedestrian route which provides an explicit representation by using Bessel functions that is convenient to get the asymptotic profiles of the solution as $\tau\rightarrow+\infty$, which is the main aim of this section.

\begin{Theorem}
 \label{teosteady}
 For any $\kappa\in\CC$, the Cauchy problem (\ref{kgdsu}), (\ref{regliss}), (\ref{initialdsu}) is well posed. The partial Fourier transform with respect to $\mathbf x$ of the solution, $\hat{u}(\tau,\pmb{\xi})={\mathcal F}_{\mathbf x}(u(\tau,.))(\pmb{\xi})$ is given by
\begin{equation}
\begin{split}
\hat{u}(\tau,\pmb{\xi})&=\frac{\pi}{2} e^{-\frac{3}{2}\tau}\left\{\left[Y'_{\nu}\left(\mid\pmb\xi\mid e^{-\tau_*}\right)J_{\nu}\left(\mid\pmb\xi\mid e^{-\tau}\right)-
J'_{\nu}\left(\mid\pmb\xi\mid e^{-\tau_*}\right)Y_{\nu}\left(\mid\pmb\xi\mid e^{-\tau}\right)\right]\mid\pmb\xi\mid e^{\frac{1}{2}\tau_*}\hat{u}_0(\pmb\xi)\right.\\
&+\left. \left[Y_{\nu}\left(\mid\pmb\xi\mid e^{-\tau_*}\right)J_{\nu}\left(\mid\pmb\xi\mid e^{-\tau}\right)-
J_{\nu}\left(\mid\pmb\xi\mid e^{-\tau_*}\right)Y_{\nu}\left(\mid\pmb\xi\mid e^{-\tau}\right)\right]e^{\frac{3}{2}\tau_*}\left(\hat{u}_1(\pmb\xi)+\frac{3}{2}\hat{u}_0(\pmb\xi)\right)\right\},
\end{split}
 \label{representation}
\end{equation}
where $\nu\in\CC$ satisfies $\Re\nu\geq 0$, $\nu^2=\frac{9}{4}-\kappa$.\\

When $\kappa\in\RR$, $s\geq 1$ the energy defined as
\begin{equation}
\mathcal{E}_{\kappa}(u,\tau):=\int \mid\partial_{\tau}u(\tau,\mathbf{x})\mid^2+e^{-2\tau}\mid\nabla_{\mathbf x}u(\tau,\mathbf{x})\mid^2+\kappa\mid u(\tau,\mathbf{x})\mid^2 d\mathbf{x}
 \label{energstead}
\end{equation}
is a decreasing function of $\tau$ and when $\kappa\geq 9/4$ we have for all $\tau\geq\tau_*$:
\begin{equation}
\int \left\vert\partial_{\tau}u(\tau,\mathbf{x})+\frac{3}{2}u(\tau,\mathbf{x})\right\vert^2 d\mathbf{x}\leq
2e^{3(\mid\tau_*\mid-\tau)}
\int\mid u_1\mid^2+\mid\nabla_{\mathbf x}u_0\mid^2+\kappa\mid u_0\mid^2d\mathbf{x},
 \label{enert}
\end{equation}
\begin{equation}
\int \mid\partial_{\tau}u(\tau,\mathbf{x})\mid^2+\kappa\mid u(\tau,\mathbf{x})\mid^2 d\mathbf{x}\leq
\min\left(\frac{3\kappa e^{-3\tau}}{\kappa-\frac{9}{4}},1\right)e^{3\mid\tau_*\mid}
\int\mid u_1\mid^2+\mid\nabla_{\mathbf x}u_0\mid^2+\kappa\mid u_0\mid^2d\mathbf{x},
 \label{enerc}
\end{equation}
\begin{equation}
\int \mid\nabla_{\mathbf x}u(\tau,\mathbf{x})\mid^2\leq 2e^{3\mid\tau_*\mid-\tau}\int\mid u_1\mid^2+\mid\nabla_{\mathbf x}u_0\mid^2+\kappa\mid u_0\mid^2d\mathbf{x}.
 \label{enerp}
\end{equation}

If $u_0$ and $u_1$ are compactly supported in $\mid\mathbf{x}\mid\leq R$, then for all $\tau\geq\tau_*$, $u(\tau,.)$ is supported in $\mid\mathbf{x}\mid\leq R+e^{-\tau_*}-e^{-\tau}.$\\

When $\kappa=0$, $u$ and $e^{2\tau}\partial_{\tau}u$ have an asymptotic profile at the time infinity : there exists $\phi\in H^{s+1}(\RR^3_{\mathbf x})$ such that
\begin{equation}
\Vert u(\tau,.)-\phi\Vert_{H^s}+\Vert e^{2\tau}\partial_{\tau}u(\tau,.)-\Delta\phi\Vert_{H^{s-1}}\longrightarrow 0,\;\tau\rightarrow+\infty,
 \label{assds}
\end{equation}
and $\phi$ is given by :
\begin{equation}
\hat{\phi}(\pmb{\xi})=\sqrt{\frac{\pi}{2}}e^{\frac{1}{2}\tau_*}\mid\pmb\xi\mid^{-\frac{1}{2}}\left\{J_{\frac{1}{2}}\left(e^{-\tau_*}\mid\pmb\xi\mid\right) \hat{u}_0(\pmb\xi)+J_{\frac{3}{2}}\left(e^{-\tau_*}\mid\pmb\xi\mid\right) e^{\tau_*} \mid\pmb\xi\mid^{-1}\hat{u}_1(\pmb\xi)\right\}.
 \label{pifi}
\end{equation}
\\

When $\kappa>0$, the solution is vanishing as $\tau\rightarrow+\infty$ : for almost all $\pmb\xi\in\RR^3$, when $0<\kappa<9/4$, $\hat{u}(\tau,\pmb\xi)=O\left(e^{(\sqrt{\frac{9}{4}-\kappa}-\frac{3}{2})\tau}\right)$, for $\kappa=9/4$, $\hat{u}(\tau,\pmb\xi)=O\left(\tau e^{-\frac{3}{2}\tau}\right)$, and when $\kappa>9/4$, $\hat{u}(\tau,\pmb\xi)=O\left( e^{-\frac{3}{2}\tau}\right)$. Moreover, given $\tau_*$, there exists $C>0$ independent of $u_0$ and $u_1$ such that for all $\tau>\tau_*$ we have :
\begin{equation}
0<\kappa<\frac{9}{4},\;
\left\{
\begin{array}{c}
\Vert u(\tau,.)\Vert_{H^{s}}+\Vert \partial_{\tau}u(\tau,.)\Vert_{H^{s-1}}\leq C e^{\max(\sqrt{\frac{9}{4}-\kappa}-\frac{3}{2},-1)\tau}
\left(\Vert u_0\Vert_{H^{s}}+\Vert u_1\Vert_{H^{s-1}}\right),\\
\Vert u(\tau,.)\Vert_{H^{s-\frac{1}{2}}}+\Vert \partial_{\tau}u(\tau,.)\Vert_{H^{s-\frac{3}{2}}}\leq C e^{(\sqrt{\frac{9}{4}-\kappa}-\frac{3}{2})\tau}\left( \Vert u_0\Vert_{H^{s}}+\Vert u_1\Vert_{H^{s-1}}\right),
\end{array}
\right.
 \label{asso}
\end{equation}

\begin{equation}
\kappa=\frac{9}{4},\;
\left\{
\begin{array}{c}
\Vert u(\tau,.)\Vert_{H^{s}}+\Vert \partial_{\tau}u(\tau,.)\Vert_{H^{s-1}}\leq C \tau e^{-\tau} \left(\Vert u_0\Vert_{H^{s}}+\Vert u_1\Vert_{H^{s-1}}\right),\\
\Vert u(\tau,.)\Vert_{H^{s-\frac{1}{2}}}+\Vert \partial_{\tau}u(\tau,.)\Vert_{H^{s-\frac{3}{2}}}\leq C \tau e^{-\frac{3}{2}\tau}\left(\Vert u_0\Vert_{H^{s}}+\Vert u_1\Vert_{H^{s-1}}\right),
\end{array}
\right.
 \label{assop}
\end{equation}

\begin{equation}
\kappa >\frac{9}{4},\;
\left\{
\begin{array}{c}
\Vert u(\tau,.)\Vert_{H^{s}}+\Vert \partial_{\tau}u(\tau,.)\Vert_{H^{s-1}}\leq C e^{-\tau}\left(\Vert u_0\Vert_{H^{s}}+\Vert u_1\Vert_{H^{s-1}}\right),\\
\Vert u(\tau,.)\Vert_{H^{s-\frac{1}{2}}}+\Vert \partial_{\tau}u(\tau,.)\Vert_{H^{s-\frac{3}{2}}}\leq C e^{-\frac{3}{2}\tau}\left(\Vert u_0\Vert_{H^{s}}+\Vert u_1\Vert_{H^{s-1}}\right),
\end{array}
\right.
 \label{assopp}
\end{equation}
\\

When $\kappa<0$, the solution can blow up at the time infinity :  there exists Schwartz functions $u_0$, $u_1$ such that the solution satisfies
\begin{equation}
\forall s\in\RR,\;\;\Vert u(\tau,.)\Vert_{H^s}\sim e^{\left(\sqrt{\frac{9}{4}-\kappa}-\frac{3}{2}\right)\tau},\;\;\tau\rightarrow+\infty.
 \label{blouz}
\end{equation}

\end{Theorem}

We make some comments on this result. A consequence of this theorem is that when $u_0,u_1\in C^{\infty}_0$, we have
$$
u(\tau,.)=O\left(e^{-\left(\frac{3}{2}-\sqrt{\frac{9}{4}-\kappa}\right)\tau}\right)\;\;in\;\;C^{\infty}_0\;\;if\;\;0\leq\kappa\neq\frac{9}{4},
$$
$$
u(\tau,.)=O\left(\tau e^{-\frac{3}{2}\tau}\right) \;\;in\;\;C^{\infty}_0\;\;if\;\;\kappa=\frac{9}{4}.
$$
This is a consequence of the estimates of decay in $H^{s-\frac{1}{2}}\times H^{s-\frac{3}{2}}$,  the Sobolev embedding, and the horizon of radius $R+e^{-\tau_*}$ when $u_j$ are supported in $\mid\mathbf{x}\mid\leq R$.
A. Vasy has established in \cite{vasy-ds} precise asymptotics in the much more large class of the asymptotically De Sitter space-times (see so \cite{baskin}), and we can deduce from his work that  when $u_0,u_1\in C^{\infty}_0$, there exists $v\in C^{\infty}_0(\RR^3)$ such that as $\tau\rightarrow\infty$ we have
$$
u(\tau,\mathbf{x})=e^{-\left(\frac{3}{2}-\sqrt{\frac{9}{4}-\kappa}\right)\tau}v(\mathbf{x})+o\left(e^{-\left(\frac{3}{2}-\sqrt{\frac{9}{4}-\kappa}\right)\tau}\right)\;\;if\;\;0\leq\kappa\neq\frac{9}{4},
$$
$$
u(\tau,\mathbf{x})=\tau e^{-\frac{3}{2}\tau}v(\mathbf{x})+O\left(e^{-\frac{3}{2}\tau}\right)\;\;if\;\;\kappa=\frac{9}{4}.
$$
The main novelty of Theorem \ref{teosteady} is the explicit formula (\ref{pifi}) for the trace at $\tau=+\infty$ when $\kappa=0$. We remark also that in this case,  there exists ``disappearing solutions'' that tends to zero as $\tau\rightarrow\infty$ since $\phi$ can be equal to zero. We note also a somewhat unexpected loss of rate of decay in $H^s\times H^{s-1}$ when $\kappa>2$. Nevertheless, these results are not optimal when $s=1$ since we can easily deduce from (\ref{enert}) with (\ref{assopp}) and (\ref{assop}) that $\Vert\partial_{\tau}u\Vert_{L^2}=O(e^{-\frac{3}{2}\tau})$ when $\kappa>9/4$ and $O(\tau e^{-\frac{3}{2}\tau})$ when $\kappa=9/4$. $\kappa=2$ is the critical mass for which the Klein-Gordon equation is conformal invariant, hence the solution can be expressed in this case from a free field on the static Einstein universe $\RR\times S^3$ (see the method by Y. Choquet-Bruhat \cite{choquet82}).  When $=9/4$ the equation belongs to the family of wave equations with variable propagation speed for which energy estimate have been established in \cite{hirosawa}. Besides, since the proof is based on the Fourier analysis and the investigation of the differential equation
$$
\left[\frac{d^2}{d\tau^2}+3\frac{d}{d\tau}+e^{-2\tau}\lambda^2+\kappa\right]u=0,\;\;\tau\in\RR,\;\;\lambda\in\RR,
$$
we could obtain a similar theorem for the Klein Gordon equation on the exponentially expanding Friedmann-Robertson-Walker universe
$$
\RR_{\tau}\times K_x,\;\;ds^2=d\tau^2-e^{2\tau}g_{ij}dx^idx^j
$$
where $(K,g)$ is any Riemanian manifold.
Parenthetically, we also could obtain several formula of products of Bessel functions, by using the propagator property of the map $(u_0,u_1)\mapsto(u(\tau),\partial_{\tau}u(\tau))$ and formulas (\ref{representation}) and (\ref{pifi}).\\

{\it Proof of Theorem \ref{teosteady}.} Let $u$ be a solution of (\ref{kgdsu}), (\ref{regliss}), (\ref{initialdsu}). To suppress the time derivative of first order, we introduce $v(\tau,\mathbf{x}):=e^{\frac{3}{2}\tau}u(\tau,\mathbf{x})$ that obeys the equation
\begin{equation}
\left[\frac{\partial^2}{\partial\tau^2}-e^{-2\tau}\Delta_{\mathbf x}+\kappa-\frac{9}{4}\right]v=0,\;\;\tau\in\RR,\;\;{\mathbf x}\in\RR^3.
 \label{kgdsv}
\end{equation}
If $v\in C^k\left(\RR_{\tau};H^s\left(\RR^3_{\mathbf x}\right)\right)$ is solution, its partial Fourier transform with respect to $\mathbf x$, $\hat{v}(\tau,\pmb{\xi})={\mathcal F}_{\mathbf x}(v(\tau,.))(\pmb{\xi})\in C^k\left(\RR_{\tau};L^2\left(\RR^3_{\pmb\xi},\left(1+\mid\pmb\xi\mid^2\right)^sd\pmb\xi\right)\right)$ satisfies
$$
\partial^2_{\tau}\hat{v}+\left(e^{-2\tau}\mid\pmb\xi\mid^2+\kappa-\frac{9}{4}\right)\hat{v}=0.
$$
We introduce $\nu\in\CC$, $\Re\nu\geq 0$, such that $\nu^2=\frac{9}{4}-\kappa$, and we put $\sigma:=e^{-\tau}\mid\pmb\xi\mid$. Then $V$ defined by $V(\sigma,\pmb\xi):=\mid\pmb\xi\mid\hat{v}(\tau,\pmb\xi)$ is solution of the Bessel equation
$$
\sigma^2\partial_{\sigma}^2V+\sigma\partial_{\sigma}V+\left(\sigma^2-\nu^2\right)V=0.
$$
With formula (10.5.2) of \cite{nist}, we deduce that
\begin{equation*}
\hat{v}(\tau,\pmb\xi)=A(\tau_*,\tau,\pmb\xi) \hat{v}(\tau_*,\pmb\xi)+B(\tau_*,\tau,\pmb\xi) \partial_{\tau}\hat{v}(\tau_*,\pmb\xi)
 \label{}
\end{equation*}
with
\begin{equation*}
\begin{split}
A(\tau_*,\tau,\pmb\xi)=&\frac{\pi}{2}\left[Y'_{\nu}\left(\mid\pmb\xi\mid e^{-\tau_*}\right)J_{\nu}\left(\mid\pmb\xi\mid e^{-\tau}\right)-
J'_{\nu}\left(\mid\pmb\xi\mid e^{-\tau_*}\right)Y_{\nu}\left(\mid\pmb\xi\mid e^{-\tau}\right)\right]\mid\pmb\xi\mid e^{-\tau_*},\\
B(\tau_*,\tau,\pmb\xi)=&\frac{\pi}{2}\left[Y_{\nu}\left(\mid\pmb\xi\mid e^{-\tau_*}\right)J_{\nu}\left(\mid\pmb\xi\mid e^{-\tau}\right)-
J_{\nu}\left(\mid\pmb\xi\mid e^{-\tau_*}\right)Y_{\nu}\left(\mid\pmb\xi\mid e^{-\tau}\right)\right],
\end{split}
 \label{}
\end{equation*}
and we conclude that $u$ satisfies (\ref{representation}).\\

Conversely, we prove that this formula allows to solve the Cauchy problem. To estimate these Fourier multipliers, we recall the asymptotics of the Bessel functions as $z\rightarrow 0^+$,
\begin{equation}
\Re\nu\geq 0,\;\;J_{\nu}(z)=\frac{1}{\Gamma(\nu+1)}\left(\frac{1}{2}z\right)^{\nu}+O(z^{\nu+2}),
 \label{Jojo}
\end{equation}
\begin{equation}
\Re\nu>0 ,\;\;Y_{\nu}(z)=-\frac{1}{\pi}\Gamma(\nu)\left(\frac{1}{2}z\right)^{-\nu}+o\left(z^{-\nu}\right),
 \label{Yoyo}
\end{equation}
\begin{equation}
Y_0(z)=\frac{2}{\pi}\ln z+O(1),
\;\;
\Re\nu=0,\;\nu\neq 0,\;Y_{\nu}(z)=O(1),
 \label{yooo}
\end{equation}
and as $0<z\rightarrow\infty$,
\begin{equation}
J_{\nu}(z)=\sqrt{\frac{2}{\pi z}}\left(\cos\left(z-\nu\frac{\pi}{2}-\frac{\pi}{4}\right)+o(1)\right),\;
Y_{\nu}(z)=\sqrt{\frac{2}{\pi z}}\left(\sin\left(z-\nu\frac{\pi}{2}-\frac{\pi}{4}\right)+o(1)\right).
 \label{bessinf}
\end{equation}
Since the derivative with respect to $z$ of  a Bessel function $\mathcal{C}_{\nu}=J_{\nu},\;Y_{\nu},$ is given by
\begin{equation}
\mathcal{C}'_{\nu}(z)=\frac{\nu}{z}\mathcal{C}_{\nu}(z)-\mathcal{C}_{\nu+1}(z)=-\frac{\nu}{z}\mathcal{C}_{\nu}(z)+\mathcal{C}_{\nu-1}(z),
 \label{derivbess}
\end{equation}
we get the asymptotics as $z\rightarrow 0$ :
\begin{equation}
\Re\nu\geq 0,\;\;\nu\neq 0,\;\;J_{\nu}'(z)=\frac{1}{2^{\nu}\Gamma(\nu)}z^{\nu-1}+O(z^{\nu+1}),\;\;J'_0(z)=-\frac{1}{2}z+O(z^2),
 \label{Jojoprime}
\end{equation}
\begin{equation}
\Re\nu>0 ,\;\;Y_{\nu}'(z)=\frac{2^{\nu}\Gamma(\nu+1)}{\pi}z^{-\nu-1}+o\left(z^{-\nu-1}\right),
 \label{Yoyoprime}
\end{equation}
\begin{equation}
\Re\nu=0,\;\;\;Y_{\nu}'(z)=O\left(\frac{1}{z}\right),
 \label{yoooprime}
\end{equation}
and as $z\rightarrow+\infty$ :
\begin{equation}
J_{\nu}'(z)=\sqrt{\frac{2}{\pi z}}\left(\cos\left(z-\nu\frac{\pi}{2}+\frac{\pi}{4}\right)+o(1)\right),\;
Y_{\nu}'(z)=\sqrt{\frac{2}{\pi z}}\left(\sin\left(z-\nu\frac{\pi}{2}+\frac{\pi}{4}\right)+o(1)\right).
 \label{bessinfprime}
\end{equation}
We estimate $A,\;\partial_{\tau}A,\;B,\;\partial_{\tau}$ by choosing the asymptotics suitable for each zone of frequency. For the high frequencies, $\mid\pmb\xi\mid\geq e^{\tau}$ or $\mid\pmb\xi\mid\geq e^{\tau_*}$ we use the behaviours  (\ref{bessinf}), (\ref{bessinfprime}), and for the low frequencies  $\mid\pmb\xi\mid\leq e^{\tau}$ or $\mid\pmb\xi\mid\leq e^{\tau_*}$, we employ the asymptotics of the Bessel functions near zero. We obtain the following bounds by investigating the Fourier multiplyers on three zones, $(I):\;\mid\xi\mid>\max\left(e^{\tau},\,e^{\tau_*}\right)$, $(II'):\;e^{\tau_*}\leq\mid\pmb\xi\mid\leq e^{\tau}$ or $(II''):\;e^{\tau}\leq\mid\pmb\xi\mid\leq e^{\tau_*}$ , and $(III):\;\mid\pmb\xi\mid<\min\left(e^{\tau},\,e^{\tau_*}\right)$. We omit the details of these long but elementary calculations that prove that there exists $C>0$ independent of $\tau,\tau_*,\,\pmb\xi$ such that :
\begin{equation}
in\;\;zone\;\;(I):\;\;\left\{
\begin{array}{c}
\mid A(\tau_*,\tau,\pmb\xi)\mid\leq Ce^{\frac{\tau-\tau_*}{2}}\leq Ce^{-\frac{\tau_*}{2}}\mid\pmb\xi\mid^{\frac{1}{2}},\\
(1+\mid\pmb\xi\mid)^{-1}\mid \partial_{\tau}A(\tau_*,\tau,\pmb\xi)\mid\leq Ce^{-\frac{\tau+\tau_*}{2}},\\
\mid B(\tau_*,\tau,\pmb\xi)\mid\leq Ce^{\frac{\tau-\tau_*}{2}}\mid\pmb\xi\mid^{-1}\leq Ce^{-\frac{\tau_*}{2}}\mid\pmb\xi\mid^{-\frac{1}{2}},\\
\mid \partial_{\tau}B(\tau_*,\tau,\pmb\xi)\mid\leq Ce^{-\frac{\tau+\tau_*}{2}},
\end{array}
\right.
 \label{I}
\end{equation}

\begin{equation}
in\;\;zone\;\;(II'):\;\;\left\{
\begin{array}{c}
if\;\;\Re\nu\geq 0,\;\nu\neq 0:\\
\mid A\mid\leq C\min\left( e^{-\frac{\tau_*}{2}}\mid\pmb\xi\mid^{\frac{1}{2}}\left(1+e^{\Re\nu(\tau-\tau_*)}\right),\;\;e^{\max(\frac{1}{2},\Re\nu) (\tau-\tau_*)}\right),\\
(1+\mid\pmb\xi\mid)^{-1}\mid \partial_{\tau}A\mid\leq C  e^{-\tau_*}\left(1+e^{\Re\nu(\tau-\tau_*)}\right),\\
(1+\mid\pmb\xi\mid)^{\frac{1}{2}}\mid B\mid\leq C(1+e^{\frac{\tau_*}{2}})\left(1+e^{\Re\nu(\tau-\tau_*)}\right),\\
(1+\mid\pmb\xi\mid)\mid B\mid\leq C(1+e^{\tau_*})e^{\max(\frac{1}{2},\Re\nu)(\tau-\tau_*)},\\
(1+\mid\pmb\xi\mid)^{-\frac{1}{2}}\mid\partial_{\tau} B\mid\leq Ce^{\frac{\tau_*}{2}}(1+e^{\Re\nu(\tau-\tau_*)}),\\
\mid\partial_{\tau} B\mid\leq Ce^{\tau_*+\max(\frac{1}{2},\Re\nu)(\tau-\tau_*)},\\
if\;\;\nu=0:\\
\mid A\mid\leq C e^{-\frac{\tau_*}{2}}\mid\pmb\xi\mid^{\frac{1}{2}}(1+\mid\tau\mid+\mid\tau_*\mid)\leq
C e^{\frac{\tau-\tau_*}{2}}(1+\mid\tau\mid+\mid\tau_*\mid),\\
(1+\mid\pmb\xi\mid)^{-1}\mid\partial_{\tau}A\mid\leq C e^{-\tau_*},\\
(1+\mid\pmb\xi\mid)^{\frac{1}{2}}\mid B\mid\leq C(1+e^{\frac{\tau_*}{2}})(1+\mid\tau\mid+\mid\tau_*\mid),\\
(1+\mid\pmb\xi\mid)\mid B\mid\leq C(1+e^{\frac{\tau+\tau_*}{2}})(1+\mid\tau\mid+\mid\tau_*\mid),\\
\mid\partial_{\tau} B\mid\leq C,
\end{array}
\right.
 \label{II'}
\end{equation}

\begin{equation}
in\;\;zone\;\;(II''):\;\;\left\{
\begin{array}{c}
if\;\;\Re\nu\geq 0,\;\nu\neq 0:\\
\mid A\mid\leq C\left(1+e^{\Re\nu(\tau_*-\tau)}\right),\\
\mid\partial_{\tau}A\mid\leq Ce^{(1+\Re\nu)(\tau_*-\tau)},\\
(1+\mid\pmb\xi\mid)^{\frac{1}{2}}\mid B\mid\leq C(1+e^{\frac{\tau}{2}})\left(1+e^{\Re\nu(\tau_*-\tau)}\right),\\
(1+\mid\pmb\xi\mid)\mid B\mid\leq C(1+e^{\frac{\tau_*+\tau}{2}})\left(1+e^{\Re\nu(\tau_*-\tau)}\right),\\
\mid \partial_{\tau}B\mid\leq C e^{\frac{\tau_*-\tau}{2}}\left(1+e^{\Re\nu(\tau_*-\tau)}\right),\\
if\;\;\nu=0:\\
\mid A\mid\leq Ce^{\tau_*-\tau},\\
\mid\partial_{\tau}A\mid\leq Ce^{\frac{\tau_*-\tau}{2}},\\
(1+\mid\pmb\xi\mid)\mid B\mid\leq Ce^{\frac{\tau_*+\tau}{2}}(1+\mid\tau\mid+\mid\tau_*\mid),\\
\mid \partial_{\tau}B\mid\leq Ce^{\frac{\tau_*-\tau}{2}}(1+\mid\tau\mid+\mid\tau_*\mid),
\end{array}
\right.
 \label{II''}
\end{equation}

\begin{equation}
in\;zone\;(III):\;
\left\{
\begin{array}{c}
if\;\;\Re\nu\geq 0,\;\nu\neq 0,\;\;
\mid A\mid,\,\mid \partial_{\tau}A\mid,\,
\mid B\mid,\,\mid \partial_{\tau}B\mid\leq Ce^{\Re\nu\mid\tau-\tau_*\mid},\\
if\;\nu=0,\;\mid A\mid,\,\mid B\mid,\,\mid\partial_{\tau} B\mid \leq C(1+\mid\tau\mid+\mid\tau_*\mid),\;\mid \partial_{\tau}A\mid\leq C
\end{array}
\right.
 \label{III}
\end{equation}

All these estimates show that :
$$A,\;\;(1+\mid\pmb{\xi}\mid)^{-1}\partial_{\tau}A,\;\;(1+\mid\pmb\xi\mid) B, \;\; \partial_{\tau}B\in L^{\infty}_{loc}\left(\RR_{\tau_*}\times\RR_{\tau}; L^{\infty}\left(\RR^3_{\pmb\xi}\right)\right).$$
Since these multiplyers are continuous with respect to $\tau,\tau_*$ for fixed $\pmb\xi$, we conclude that given $u_0\in H^s(\RR^3)$, $u_1\in H^{s-1}(\RR^3)$, the fonction $u$ defined by the formula (\ref{representation}), is a solution of (\ref{kgdsu}), (\ref{regliss}), (\ref{initialdsu}), and the map $(u_0,u_1)\longmapsto (u(\tau,.),\partial_{\tau}u(\tau,.))$ defines un propagator  $C^0 \left(\RR_{\tau_*,\tau}^2;\mathcal{L}\left(H^s\times H^{s-1}\right)-weak\right)$, {\it i.e} the Cauchy problem is well-posed.\\

To prove the energy is decreasing, it is sufficient to consider the case $u_0\in H^2$, $u_1\in H^1$ for which $u\in C^1\left(\RR_{\tau};H^1(\RR^3_{\mathbf x})\right)$ and note that
$$
\frac{d\mathcal{E}_{\kappa}}{d\tau}=-2\int_{\RR^3}3\mid\partial_{\tau}u\mid^2+e^{-2\tau}\mid\nabla_{\mathbf x}u\mid^2d\mathbf{x}\leq 0.
$$
Then we conclude by an argument of density. By a similar way we can prove that the energy of $v$, defined by
$$
\int\mid\partial_{\tau}v(\tau,\mathbf{x})\mid^2+e^{-2\tau}\mid\nabla_{\mathbf x}v(\tau,\mathbf{x})\mid^2+\left(\kappa-\frac{9}{4}\right)\mid v(\tau,\mathbf{x})\mid^2d\mathbf{x}
$$
is decreasing. We remark that we have
$
\mid u\partial_{\tau}u\mid\leq \frac{1}{2\sqrt{\kappa}}(\mid\partial_{\tau}u\mid^2+\kappa\mid u\mid^2),
$
hence we deduce that for $\kappa\geq9/4$
\begin{equation*}
\begin{split}
e^{3\tau}\left(1-\frac{3}{2\sqrt{\kappa}}\right)\int\mid&\partial_{\tau}u(\tau,\mathbf{x})\mid^2+\kappa\mid u(\tau,\mathbf{x})\mid^2d\mathbf{x}+
e^{\tau}\int\mid\nabla_{\mathbf x}u(\tau,\mathbf{x})\mid^2d\mathbf{x}\\
&\leq
\int e^{3\tau}\mid\partial_{\tau}u(\tau,\mathbf{x})+\frac{3}{2}u(\tau,\mathbf{x})\mid^2+e^{\tau}\mid\nabla_{\mathbf x}u(\tau,\mathbf{x})\mid^2+e^{3\tau}\left(\kappa-\frac{9}{4}\right)\mid u(\tau,\mathbf{x})\mid^2d\mathbf{x}\\
&\leq
\int e^{3\tau_*}\mid u_1+\frac{3}{2}u_0\mid^2+e^{\tau_*}\mid\nabla_{\mathbf x}u_0\mid^2+e^{3\tau_*}\left(\kappa-\frac{9}{4}\right)\mid u_0\mid^2d\mathbf{x}\\
&\leq e^{3\mid\tau_*\mid}\left(1+\frac{3}{2\sqrt{\kappa}}\right)\int\mid u_1\mid^2+\kappa\mid u_0\mid^2d\mathbf{x}+e^{\mid\tau_*\mid}\int
\mid\nabla_{\mathbf x}u_0\mid^2d\mathbf{x}\\
&\leq 2e^{3\mid\tau_*\mid}\int\mid u_1\mid^2+\kappa\mid u_0\mid^2+\mid\nabla_{\mathbf x}u_0\mid^2d\mathbf{x}.
\end{split}
\end{equation*}
We also deduce from the decay of $\mathcal{E}_{\kappa}$ that for any $\tau\geq\tau_*$
$$
\int\mid\partial_{\tau}u(\tau,\mathbf{x})\mid^2+\kappa\mid u(\tau,\mathbf{x})\mid^2d\mathbf{x}\leq
e^{2\mid\tau_*\mid}\int\mid u_1\mid^2+\mid\nabla_{\mathbf x}u_0\mid^2+\kappa\mid u_0\mid^2d\mathbf{x}.
$$
Now (\ref{enert}),  (\ref{enerc}) and (\ref{enerp}) follow easily from these inequalities.\\


To establish the result of finite velocity propagation, it is sufficient to consider the case of real valued compactly supported smooth initial data, $u_0,u_1\in C^{\infty}_0\left(\RR^3_{\mathbf x}\right)$, and we show that if $u_0=u_1=0$ on $\mid\mathbf{x}\mid>R_0$, then for $\tau\geq\tau_*$, $v(\tau,\mathbf{x})=0$ on $\mid\mathbf{x}\mid>R_0+e^{-\tau_*}-e^{-\tau}$. First we suppose $\kappa=\frac{9}{4}$, hence the $C^{\infty}$-solutions of (\ref{kgdsv}) satisfy 
$$
\nabla_{\tau,\mathbf{x}}\cdot\left(\mid\partial_{\tau}v\mid^2+e^{-2\tau}\mid\nabla_{\mathbf x}v\mid^2,-2e^{-2\tau}\partial_{\tau}v\nabla_{\mathbf x}v\right)=-2e^{-2\tau}\mid\nabla_{\mathbf x}v\mid^2.
$$
Given $R>0$, $T>\tau_*$, we integrate this divergence on the domain $\left\{(\tau,\mathbf{x});\;\mid\mathbf{x}\mid< R+e^{-\tau}-e^{-\tau_*},\;\tau_*<\tau<T\right\}$ and since 
$$
e^{-\tau}\mid\partial_{\tau}v\mid^2+e^{-3\tau}\mid\nabla_{\mathbf x}v\mid^2-2e^{-2\tau}\partial_{\tau}v\nabla_{\mathbf x}v\cdot\frac{\mathbf x}{\mid\mathbf x\mid}\geq \left(e^{-\frac{\tau}{2}}\mid\partial_{\tau}v\mid-e^{-\frac{3\tau}{3}}\mid\nabla_{\mathbf x}v\mid\right)^2,
$$
we obtain an estimate of the local energy :
\begin{equation*}
\mathcal{E}\left(v,T,R+e^{-T}-e^{-\tau_*}\right)\leq \mathcal{E}\left(v,\tau_*,R\right),
 \label{}
\end{equation*}
where the local energy is defined as
\begin{equation*}
\mathcal{E}\left(v,\tau,R\right):=\int_{\mid\mathbf{x}\mid<R}\mid\partial_{\tau}v(\tau,\mathbf{x})\mid^2+e^{-2\tau}\mid\nabla_{\mathbf x}v(\tau,\mathbf{x})\mid^2\;d\mathbf{x}.
 \label{}
\end{equation*}
Now we fix $\mathbf{x}_0$ such that $\mid\mathbf{x}_0\mid>R-e^{-T}+e^{-\tau_*}$, and we apply the previous estimate to $w_{\mathbf{x}_0}(\tau,\mathbf{x}):=v(\tau,\mathbf{x}_0+\mathbf{x})$ that is solution of (\ref{kgdsv}) again, to get :
$$
\mathcal{E}\left(w_{\mathbf{x}_0},\tau,\mid\mathbf{x}_0\mid-R_0+e^{-\tau}-e^{-\tau_*}\right)\leq \mathcal{E}\left(w_{\mathbf{x}_0},\tau_*,\mid\mathbf{x}_0\mid-R_0\right)=0,\;\;\tau_*\leq\tau\leq T.
$$
We deduce that $\partial_{\tau}v(\tau,\mathbf{x}_0)=0$ for all $\tau\in[\tau_*,T]$ and since $v(\tau_*,\mathbf{x}_0)=0$, we conclude that $v(T,\mathbf{x}_0)=0$. Hence we have proved that for any $\tau,\tau_*$, $\tau\geq\tau_*$,  the propagator $U_{0}(\tau,\tau_*)$ that associates $(v(\tau,.),\partial_{\tau}v(\tau,.))$, where $v$ is solution of (\ref{kgdsv}) with $\kappa=\frac{9}{4}$, to the initial data $(v(\tau_*,.),\partial_{\tau_*}v(\tau,.))$, is a bounded operator from $H^s_R\times H^{s-1}_R$ to $H^s_{R+e^{-\tau_*}-e^{-\tau}}\times H^{s-1}_{R+e^{-\tau_*}-e^{-\tau}}$, where $H^s_R$ is the set of distributions in $H^s(\RR^3)$ of which the support is included in $\mid\mathbf{x}\mid\leq R$. Finally for all $\kappa$, we note that the solution $v$ of (\ref{kgdsv}) satisfies
$$
\left(\begin{array}{c}
v(\tau)\\
\partial_{\tau}v(\tau)
\end{array}\right)
=
U_0(\tau,\tau_*)
\left(\begin{array}{c}
v(\tau_*)\\
\partial_{\tau}v(\tau_*)
\end{array}\right)
+
\int_{\tau_*}^{\tau}
U_0(\tau,\sigma)
\left(\begin{array}{c}
0\\
\left(\frac{9}{4}-\kappa\right)v(\sigma)
\end{array}\right)d\sigma
$$
and this equation can be easily solved by the Picard iterates,
$$
\left(\begin{array}{c}
v(\tau)\\
\partial_{\tau}v(\tau)
\end{array}\right)
=
\sum_{n=0}^{\infty}
\left(\begin{array}{c}
v_n(\tau)\\
v'_n(\tau)
\end{array}\right)
$$
where
$$
\left(\begin{array}{c}
v_0(\tau)\\
v'_0(\tau)
\end{array}\right)
:=U_0(\tau,\tau_*)
\left(\begin{array}{c}
v(\tau_*)\\
\partial_{\tau}v(\tau_*)
\end{array}\right),
\;\;
\left(\begin{array}{c}
v_n(\tau)\\
v'_n(\tau)
\end{array}\right)
:=
\int_{\tau_*}^{\tau}
U_0(\tau,\sigma)
\left(\begin{array}{c}
0\\
\left(\frac{9}{4}-\kappa\right)v_{n-1}(\sigma)
\end{array}\right)d\sigma,\;\;1\leq n,
$$
because we can show by recurrence that for $\tau\in[\tau_*,T]$, we have with $M:=\sup_{\tau\in[\tau_*,T]}\left\Vert U_0(\tau,\tau_*)\right\Vert_{\mathcal{L}(H^s\times H^{s-1})}$
$$
\left\Vert\left(\begin{array}{c}
v_n(\tau)\\
v'_n(\tau)
\end{array}\right)
\right\Vert_{H^s\times H^{s-1}}
\leq \frac{\left(M\mid\frac{9}{4}-\kappa\mid\mid\tau-\tau_*\mid\right)^n}{n!}M \left\Vert\left(\begin{array}{c}
v(\tau_*)\\
v'(\tau_*)
\end{array}\right)
\right\Vert_{H^s\times H^{s-1}}.
$$
If we assume that $v_{n-1}(\sigma,.)$ is supported in $\mid\mathbf{x}\mid\leq R+e^{-\tau_*}-e^{-\sigma}$, the previous result for $U_0$ assures that $v_n(\tau,.)$ is supported in $\mid\mathbf{x}\mid\leq R+e^{-\tau_*}-e^{-\tau}$ and the proof is complete.\\

To prove the existence of the asymptotic profile $\phi$ and the results of decay, we estimate the Fourier multipliers as $\tau\rightarrow\infty$, $\tau_*$ and $\pmb\xi$ being fixed :
\begin{equation*}
e^{-\frac{3}{2}\tau}A(\tau_*,\tau,\pmb\xi)=
\left\{
\begin{array}{c}
O\left(\tau e^{-\frac{3}{2}\tau}\right)\;\;if\;\;\nu=0,\\
O\left(e^{\left(\Re\nu-\frac{3}{2}\right)\tau}\right)\;\;if\;\;0\leq\Re\nu<\frac{3}{2},\;\;\nu\neq 0,\\
\sqrt{\frac{\pi}{2}}e^{-\tau_*}J_{\frac{3}{2}}'\left(e^{-\tau_*}\mid\pmb\xi\mid\right) \mid\pmb\xi\mid^{-\frac{1}{2}}+o(1),\;\;\nu=\frac{3}{2},
\end{array}\right.
 \label{}
\end{equation*}

\begin{equation*}
e^{-\frac{3}{2}\tau}B(\tau_*,\tau,\pmb\xi)=
\left\{
\begin{array}{c}
O\left(\tau e^{-\frac{3}{2}\tau}\right)\;\;if\;\;\nu=0,\\
O\left(e^{\left(\Re\nu-\frac{3}{2}\right)\tau}\right)\;\;if\;\;0\leq\Re\nu<\frac{3}{2},\;\;\nu\neq 0,\\
\sqrt{\frac{\pi}{2}}J_{\frac{3}{2}}\left(e^{-\tau_*}\mid\pmb\xi\mid\right) \mid\pmb\xi\mid^{-\frac{3}{2}}+o(1),\;\;\nu=\frac{3}{2},
\end{array}\right.
 \label{}
\end{equation*}

\begin{equation*}
e^{-\frac{3}{2}\tau}\partial_{\tau}A(\tau_*,\tau,\pmb\xi)=
\left\{
\begin{array}{c}
O\left(e^{\left(\Re\nu-\frac{3}{2}\right)\tau}\right)\;\;if\;\;0\leq\Re\nu<\frac{3}{2},\\
\frac{3}{2}\sqrt{\frac{\pi}{2}}e^{-\tau_*}J_{\frac{3}{2}}'\left(e^{-\tau_*}\mid\pmb\xi\mid\right) \mid\pmb\xi\mid^{-\frac{1}{2}}+o(1),\;\;\nu=\frac{3}{2},
\end{array}\right.
 \label{}
\end{equation*}

\begin{equation*}
e^{-\frac{3}{2}\tau}\partial_{\tau}B(\tau_*,\tau,\pmb\xi)=
\left\{
\begin{array}{c}
O\left(e^{\left(\Re\nu-\frac{3}{2}\right)\tau}\right)\;\;if\;\;0\leq\Re\nu<\frac{3}{2},\\
\frac{3}{2}\sqrt{\frac{\pi}{2}}J_{\frac{3}{2}}\left(e^{-\tau_*}\mid\pmb\xi\mid\right) \mid\pmb\xi\mid^{-\frac{3}{2}}+o(1),\;\;\nu=\frac{3}{2},
\end{array}\right.
 \label{}
\end{equation*}

Since ${\mathcal C}_{\frac{3}{2}}'(z)+\frac{3}{2z}{\mathcal C}_{\frac{3}{2}}(z)={\mathcal C}_{\frac{1}{2}}(z)$, ${\mathcal C}=J,Y$, we also have when $\nu=\frac{3}{2}$ :
\begin{equation*}
\partial_{\tau}A-\frac{3}{2}A=-\frac{\pi}{2}\mid\pmb\xi\mid^2 e^{-\tau-\tau_*}\left[Y'_{\frac{3}{2}}(\mid\pmb\xi\mid e^{-\tau_*})J_{\frac{1}{2}}(\mid\pmb\xi\mid e^{-\tau})-J'_{\frac{3}{2}}(\mid\pmb\xi\mid e^{-\tau_*})Y_{\frac{1}{2}}(\mid\pmb\xi\mid e^{-\tau})\right],
 \label{}
\end{equation*}
\begin{equation*}
\partial_{\tau}B-\frac{3}{2}B=-\frac{\pi}{2}\mid\pmb\xi\mid e^{-\tau}\left[Y_{\frac{3}{2}}(\mid\pmb\xi\mid e^{-\tau_*})J_{\frac{1}{2}}(\mid\pmb\xi\mid e^{-\tau})-J_{\frac{3}{2}}(\mid\pmb\xi\mid e^{-\tau_*})Y_{\frac{1}{2}}(\mid\pmb\xi\mid e^{-\tau})\right],
 \label{}
\end{equation*}
and we get in this case :
\begin{equation*}
e^{\frac{1}{2}\tau}(\partial_{\tau}A-\frac{3}{2}A)=-\sqrt{\frac{\pi}{2}}\mid\pmb\xi\mid^{\frac{3}{2}}e^{-\tau_*} J'_{\frac{3}{2}}(e^{-\tau_*}\mid\pmb\xi\mid)+O(e^{-\tau}),
 \label{}
\end{equation*}
\begin{equation*}
e^{\frac{1}{2}\tau}(\partial_{\tau}B-\frac{3}{2}B)=-\sqrt{\frac{\pi}{2}}\mid\pmb\xi\mid^{\frac{1}{2}}J_{\frac{3}{2}}(e^{-\tau_*}\mid\pmb\xi\mid)+O(e^{-\tau}).
 \label{}
\end{equation*}

We deduce that, as $\tau\rightarrow+\infty$, for almost all $\pmb\xi$, $\hat{u}(\tau,\pmb\xi)$ tends to $\hat{\phi}(\pmb\xi)$ given by (\ref{pifi}) when $\kappa=0$, $\hat{u}(\tau,\pmb\xi)$ and $\partial_{\tau}\hat{u}(\tau,\pmb\xi)$ are $o(1)$ when $\kappa >0$, and $e^{2\tau}\partial_{\tau}\hat{u}(\tau,\pmb\xi)+\mid\pmb\xi\mid^2\hat{\phi}(\pmb\xi)=o(1)$ when $\kappa=0$. Moreover, since  $J_{\frac{3}{2}}\left(e^{-\tau_*}\mid\pmb\xi\mid\right)$ and $J_{\frac{3}{2}}'\left(e^{-\tau_*}\mid\pmb\xi\mid\right)$ are $O\left(\mid\pmb\xi\mid^{-\frac{1}{2}}\right)$ as $\mid\pmb\xi\mid\rightarrow\infty$, we get that $\phi\in H^{s+1}$.
To obtain the asymptotic behaviours (\ref{assds}) of $u(\tau,.)$, $e^{2\tau}\partial_{\tau}u(\tau,.)$ when $\kappa=0$ in the Sobolev spaces, it is sufficient to prove that $\tau_*$ being fixed, we have
\begin{equation}
\sup_{\tau\in[\tau_*,\infty[}\sup_{\pmb\xi\in\RR^3}e^{-\frac{3}{2}\tau}\left(\mid A\mid+(1+\mid\pmb\xi\mid)^{-1}\mid \partial_{\tau}A\mid+ (1+\mid\pmb\xi\mid)\mid B\mid +\mid \partial_{\tau}B\mid\right)<\infty,\;\;\nu=\frac{3}{2},
 \label{domingun}
\end{equation}
\begin{equation}
\sup_{\tau\in[\tau_*,\infty[}\sup_{\pmb\xi\in\RR^3}e^{\frac{1}{2}\tau}\left((1+\mid\pmb\xi\mid)^{-1}\mid \partial_{\tau}A-\frac{3}{2}A\mid+\mid \partial_{\tau}B-\frac{3}{2}B\mid\right)<\infty,\;\nu=\frac{3}{2},
 \label{domindeu}
\end{equation}
and we achieve with the dominated convergence theorem. (\ref{domingun}) is a direct consequence of the estimates (\ref{I}),  (\ref{II'}),  (\ref{III}) when $\nu=3/2$. To get (\ref{domindeu}), we estimate $\partial_{\tau}A-\frac{3}{2}A$ and  $\partial_{\tau}B-\frac{3}{2}B$ in the three zones $(I)$, $(II')$, $(III)$ as following :
\begin{equation*}
in\;\;zone\;\;(I):
\left\{
\begin{array}{c}
(1+\mid\pmb\xi\mid)^{-1}\mid \partial_{\tau}A-\frac{3}{2}A\mid\leq Ce^{-\frac{\tau+\tau_*}{2}},\\
\mid \partial_{\tau}B-\frac{3}{2}B\mid\leq Ce^{-\frac{\tau-\tau_*}{2}},
\end{array}
\right.
 \label{}
\end{equation*}

\begin{equation*}
in\;\;zone\;\;(II'):
\left\{
\begin{array}{c}
(1+\mid\pmb\xi\mid)^{-1}\mid \partial_{\tau}A-\frac{3}{2}A\mid\leq Ce^{-\frac{\tau+\tau_*}{2}},\\
\mid \partial_{\tau}B-\frac{3}{2}B\mid\leq Ce^{-\frac{\tau-\tau_*}{2}},
\end{array}
\right.
 \label{}
\end{equation*}

\begin{equation*}
in\;\;zone\;\;(III):
\left\{
\begin{array}{c}
\mid \partial_{\tau}A-\frac{3}{2}A\mid,\;\;\mid \partial_{\tau}B-\frac{3}{2}B\mid\leq Ce^{-\frac{\tau-\tau_*}{2}}.
\end{array}
\right.
 \label{}
\end{equation*}

To establish the decay results (\ref{asso}), (\ref{assop}), (\ref{assopp}), we remark that the estimates (\ref{I}),  (\ref{II'}), (\ref{II''}), (\ref{III}) assure that given $\tau_*$ there exists $C>0$ such that for any $\tau\geq\tau_*$ we have for all $\pmb\xi\in \RR^3$ :
\begin{equation*}
\left\{
\begin{array}{c}
when\;\;\Re\nu\geq 0,\;\; and\;\; \nu\neq 0:\\
(1+\mid\pmb\xi\mid)^{-\frac{1}{2}}\mid A\mid,\;(1+\mid\pmb\xi\mid)^{-\frac{3}{2}}\mid \partial_{\tau}A\mid,\;(1+\mid\pmb\xi\mid)^{\frac{1}{2}}\mid B\mid,\;(1+\mid\pmb\xi\mid)^{-\frac{1}{2}}\mid\partial_{\tau} B\mid\leq Ce^{\Re\nu\tau},\\
\mid A\mid,\;(1+\mid\pmb\xi\mid)^{-1}\mid\partial_{\tau}A\mid,\;(1+\mid\pmb\xi\mid)\mid B\mid,\;\mid\partial_{\tau}B\mid\leq Ce^{\max(\frac{1}{2},\Re\nu)\tau},\\
when\;\;\nu=0:\\
(1+\mid\pmb\xi\mid)^{-\frac{1}{2}}\mid A\mid,\;(1+\mid\pmb\xi\mid)^{-\frac{3}{2}}\mid \partial_{\tau}A\mid,\;(1+\mid\pmb\xi\mid)^{\frac{1}{2}}\mid B\mid,\;(1+\mid\pmb\xi\mid)^{-\frac{1}{2}}\mid\partial_{\tau} B\mid\leq C(1+\mid\tau\mid),\\
\mid A\mid,\;(1+\mid\pmb\xi\mid)^{-1}\mid\partial_{\tau}A\mid,\;(1+\mid\pmb\xi\mid)\mid B\mid,\;\mid\partial_{\tau}B\mid\leq Ce^{\frac{\tau}{2}}(1+\mid\tau\mid),
\end{array}
\right.
 \label{}
\end{equation*}
and we conclude with the formula $\hat{u}(\tau)=e^{-\frac{3}{2}(\tau-\tau_*)}\left(A\hat{u}_0+B(\hat{u}_1+\frac{3}{2}\hat{u}_0)\right)$.

Finally to prove the blow-up when $\kappa<0$, we choose $u_0=0$ and $\hat{u}_1\in C^{\infty}_0(\RR^3)\setminus\{0\}$, and we deduce from (\ref{Jojo}) and (\ref{Yoyo}) that
$$
\sup_{\pmb\xi}\left\vert\hat{u}(\tau,\pmb\xi)-\frac{1}{2\nu}e^{\left(\nu-\frac{3}{2}\right)\tau}\hat{u}_1(\pmb\xi)\right\vert=o\left(e^{\left(\nu-\frac{3}{2}\right)\tau}\right),
$$
and this estimate assures (\ref{blouz}).
\fin



\section{The Kaluza-Klein Tower}
In this section we show that any finite energy solution of (\ref{eqq}) and (\ref{condlim}) can be expressed as a superposition of Klein-Gordon fields propagating in the Steady State universe, called {\it Kaluza-Klein tower}. The key of the proof is a complete spectral analysis of the operator $L_c$ that we perform now. In the sequel $P^{-\mu}_{\nu}$ and ${\pmb Q}^{\mu}_{\nu}$  are the associated Legendre functions of first and second kind (see {\it e.g} \cite{nist}).

\begin{Proposition}
 \label{}
 For all $c\in\RR\cup\{\infty\}$ the spectrum of the self-adjoint operator $L_c$ has the following from :
\begin{equation*}
\sigma_{ac}(L_c)=[9/4,\infty[,\;\;\sigma_{sc}(L_c)=\emptyset,
 \label{}
\end{equation*}
$\sigma_{p}(L_c)$ is a finite subset included in $]-\infty,9/4[$, and the eigenvalues are the solutions $\lambda<\frac{9}{4}$ of the transcendental equation
\begin{equation}
\left\{
\begin{array}{c}
\begin{split}
\left(c\sqrt{1-\alpha^2}-2+\sqrt{M^2+4}\right)&P^{-\sqrt{\frac{9}{4}-\lambda}}_{-\frac{1}{2}+\sqrt{M^2+4}}(-\alpha^{-1})\\
+\alpha&\left(-\frac{1}{2}+\sqrt{M^2+4}-\sqrt{\frac{9}{4}-\lambda}\right) P^{-\sqrt{\frac{9}{4}-\lambda}}_{-\frac{3}{2}+\sqrt{M^2+4}}(-\alpha^{-1})=0\;when\;c\in\RR,\\
P^{-\sqrt{\frac{9}{4}-\lambda}}_{-\frac{1}{2}+\sqrt{M^2+4}}(-\alpha^{-1})=0\;when&\;c=\infty,
\end{split}
\end{array}
\right.
 \label{transcendante}
\end{equation}
and the corresponding eigenfunctions are given by :
\begin{equation}
w(\rho;\lambda)=\gamma (\sinh\rho)^{-\frac{3}{2}}P^{-\sqrt{\frac{9}{4}-\lambda}}_{-\frac{1}{2}+\sqrt{M^2+4}}(\cosh\rho),\;\gamma\in\CC.
 \label{foncprop}
\end{equation}
For $c=M=0$, we have
\begin{equation}
\sigma_p(L_0)=\{0\}.
 \label{pointspectrum}
\end{equation}
For $c=\infty$, we have
\begin{equation*}
\forall \alpha\in]-1,0[,\;\;\sigma_p(L_{\infty})\cap\left]-\infty,\frac{5}{4}\right[=\emptyset,
 \label{}
\end{equation*}
\begin{equation*}
\forall\lambda\in\left[\frac{5}{4},\frac{9}{4}\right[,\;\forall M\geq0,\;\exists !\;\alpha\in]-1,0[,\;\lambda\in\sigma_p(L_{\infty}).
 \label{}
\end{equation*}
\end{Proposition}

{\it Proof.}
It will be very convenient to use the Liouville normal form of operator $L$. We replace $\rho$ by a new space variable $y$ defined by
\begin{equation*}
y:=\log\left(\frac{e^{\rho}+1}{e^{\rho}-1}\right),\;y_0:=\log\left(\frac{e^{\rho_0}+1}{e^{\rho_0}-1}\right)=\log\left(\frac{1-\alpha+\sqrt{1-\alpha^2}}{1+\alpha+\sqrt{1-\alpha^2}}\right),
 \label{}
\end{equation*}
and we introduce a change of function :
\begin{equation*}
F:u\mapsto \tilde{v},\;\;\tilde{v}(y):=\left(\sinh\rho\right)^{\frac{3}{2}}u(\rho),\;y\in[y_0,\infty[,\;\rho\in]0,\rho_0].
 \label{}
\end{equation*}
We easily check that the map $F$ is an isometry from $H$ onto $L^2(y_0,\infty)$, and an isomorphism from $H_1$ onto $H^1(y_0,\infty)$, from $\dot{H}_1$ onto $H^1_0(y_0,\infty)$, from $D_{max}$ onto $H^2(y_0,\infty)$. As a consequence of the Sobolev embedding, we get asymptotics of $u$ near $\rho=0$:
\begin{equation}
u\in H_1\Rightarrow u(\rho)=o\left((\sinh \rho)^{-\frac{3}{2}}\right),\;\;u\in D_{max}\Rightarrow u'(\rho)=o\left((\sinh \rho)^{-\frac{5}{2}}\right),\;\;\rho\rightarrow 0.
 \label{ureau}
\end{equation}

The main interest of this coordinate is that the self-adjoint operator $L_{c}$ is unitarily equivalent to a Schr\"odinger type operator $\tilde{L}_{c}$ ,
\begin{equation*}
L_{c}=F^{-1}\tilde{L}_{c}F,\;\;
\tilde{L}_{c}:=-\frac{d^2}{dy^2}+\left(\frac{9}{4}+\frac{M^2+\frac{15}{4}}{\sinh^2y}\right),
 \label{}
\end{equation*}
endowed with the domain
\begin{equation*}
c\in\RR,\;\tilde{D}_{c}:=\left\{\tilde{v}\in H^2(y_0,\infty);\;\tilde{v}'(y_0)=\left(\frac{3}{2\alpha}-\frac{c \sqrt{1-\alpha^2}}{\alpha}\right)\tilde{v}(y_0)\right\},\;\;\tilde{D}_{\infty}:=H^2\cap H^1_0(y_0,\infty).
 \label{}
\end{equation*}
Hence the study of the spectral properties of $L_c$ is reduced to the investigation of this Schr\"odinger operator on the half-line with a positive short range potential, and this case is very well known. The Weidmann theorem (see {\it e.g.} \cite{teschl}, Theorem 9.38) assures that $\sigma_{ac}(\tilde{L_c})=[9/4,\infty[$, $\sigma_{sc}(\tilde{L}_c)=\emptyset$, and $\sigma_{pp}(\tilde{L}_c)$ is a finite subset included in $]-\infty,9/4]$. To determine the point spectrum, we introduce a new coordinate $x$.
Given a function $u$ defined on $J$,  we put
\begin{equation*}
x:=\cosh\rho\in]1,-\frac{1}{\alpha}],\;\;v(x):=\left(\sinh\rho\right)^{\frac{3}{2}}u(\rho),\;\;\rho\in]0,\rho_0],
 \label{}
\end{equation*}
and we can easily check that for any $\lambda\in\CC$ :
\begin{equation*}
Lu=\lambda u,\;\;\rho\in J\Leftrightarrow (1-x^2)v''-2xv'+\left(M^2+\frac{15}{4}-\frac{\frac{9}{4}-\lambda}{1-x^2}\right)v=0,\;\;x\in J':=]1,-\alpha^{-1}[,
 \label{}
\end{equation*}
\begin{equation*}
u\in H\Leftrightarrow v\in L^2\left(J',\frac{1}{x^2-1}dx\right),
 \label{}
\end{equation*}
\begin{equation}
\frac{du}{d\rho}(\rho_0)+cu(\rho_0)=0\Leftrightarrow\frac{dv}{dx}\left(-\alpha^{-1}\right)=\frac{\alpha}{\sqrt{1-\alpha^2}}\left(c-\frac{3}{2\sqrt{1-\alpha^2}}\right)v \left(-\alpha^{-1}\right).
 \label{clifi}
\end{equation}
We recognize the associated Legendre equation, hence we introduce complex parameters $\mu$ and $\nu$ such that
\begin{equation}
\nu(\nu+1)=M^2+\frac{15}{4},\;\Re\nu\geq\frac{3}{2},\;\;\mu^2=\frac{9}{4}-\lambda,\;\Re\mu\geq 0.
 \label{munu}
\end{equation}
We know that for $\Re\mu\geq 0$, $\Re\nu\geq-\frac{1}{2}$, two independent solutions, which are real valued when $\mu$ and $\nu$ are real, are the first kind and second kind Legendre functions $P^{-\mu}_{\nu}$ and $\pmb{Q}^{\mu}_{\nu}$. Hence we can calculate the expression of $v$, given $v(x_*)$ and $v'(x_*)$ for some $x_*\in]1,-\alpha^{-1}]$,  by using the Wronskian formula (14.2.8) of \cite{nist} :
\begin{equation}
\begin{split}
v(x)=\Gamma(\mu+\nu+1)(x_*^2-1)&\left\{\left[\pmb{Q}^{\mu}_{\nu}(x_*)v'(x_*)-\left(\pmb{Q}^{\mu}_{\nu}\right)'(x_*)v(x_*)\right]P^{-\mu}_{\nu}(x)\right.\\
&\left.+\left[\left(P^{-\mu}_{\nu}\right)'(x_*)v(x_*)-P^{-\mu}_{\nu}(x_*)v'(x_*)\right]\pmb{Q}^{\mu}_{\nu}(x)\right\}.
\end{split}
 \label{ororv}
\end{equation}
We recall that, for $\Re\mu\geq 0$, $\Re\nu\geq 0$, the asymptotic behaviours as $x$ tends to $1$ are given by (see formula (14.3.9), (14.8.9), (14.8.11) and the formula of connection (14.9.15) in \cite{nist}) :
\begin{equation}
P^{-\mu}_{\nu}(x)= \left(\frac{x-1}{x+1}\right)^{\frac{\mu}{2}}\left[\frac{1}{\Gamma(1+\mu)}+O(1-x)\right],
 \label{asympp}
\end{equation}
\begin{equation}
{\pmb Q}^{\mu}_{\nu}(x)\sim \frac{\Gamma(\mu)}{2\Gamma(\mu+\nu+1)}\left(\frac{2}{x-1}\right)^{\frac{\mu}{2}},\;\;\Re\mu>0,
 \label{asympq}
\end{equation}
\begin{equation}
\begin{split}
{\pmb Q}^{\mu}_{\nu}(x)=\frac{\pi}{2\sin(\mu\pi)}&\left[\left(\frac{x-1}{x+1}\right)^{-\frac{\mu}{2}}\frac{1}{\Gamma(1-\mu)\Gamma(\nu+\mu+1)}\right.\\
&\left.-\left(\frac{x-1}{x+1}\right)^{\frac{\mu}{2}}\frac{1}{\Gamma(1+\mu)\Gamma(\nu-\mu+1)}\right]+O(x-1)
,\;\Re\mu=0,\;\mu\neq 0,
\end{split}
 \label{asympq0}
\end{equation}
\begin{equation*}
{\pmb Q}^{0}_{\nu}(x)=-\frac{1}{2\Gamma(\nu+1)}\log(x-1)+O(1).
 \label{}
\end{equation*}
Hence we conclude that $\varphi\in H$ if and only if  $\Re\mu>0$ and
\begin{equation*}
\left\{
\begin{array}{c}
v(x)=\Gamma(\mu+\nu+1)(x_*^2-1)\left[\pmb{Q}^{\mu}_{\nu}(x_*)v'(x_*)-\left(\pmb{Q}^{\mu}_{\nu}\right)'(x_*)v(x_*)\right]P^{-\mu}_{\nu}(x),\\
\left(P^{-\mu}_{\nu}\right)'(x_*)v(x_*)=P^{-\mu}_{\nu}(x_*)v'(x_*).
\end{array}
\right.
 \label{}
\end{equation*}
We deduce that $\lambda\in\RR$ is an eigenvalue of $L_c$ iff $\lambda<9/4$ and, when $c\in\RR$,
\begin{equation}
\left(P^{-\sqrt{\frac{9}{4}-\lambda}}_{-\frac{1}{2}+\sqrt{M^2+4}}\right)'(-\alpha^{-1})=\frac{\alpha}{\sqrt{1-\alpha^2}}\left(c-\frac{3}{2\sqrt{1-\alpha^2}}\right)P^{-\sqrt{\frac{9}{4}-\lambda}}_{-\frac{1}{2}+\sqrt{M^2+4}}(-\alpha^{-1}),
 \label{ektr}
\end{equation}
\begin{equation}
\pmb{Q}^{\sqrt{\frac{9}{4}-\lambda}}_{-\frac{1}{2}+\sqrt{M^2+4}}(-\alpha^{-1})\frac{\alpha}{\sqrt{1-\alpha^2}}\left(c-\frac{3}{2\sqrt{1-\alpha^2}}\right)-\left(\pmb{Q}^{\sqrt{\frac{9}{4}-\lambda}}_{-\frac{1}{2}+\sqrt{M^2+4}}\right)'(-\alpha^{-1})\neq 0,
 \label{ektrq}
\end{equation}
and when $c=\infty$,
\begin{equation}
P^{-\sqrt{\frac{9}{4}-\lambda}}_{-\frac{1}{2}+\sqrt{M^2+4}}(-\alpha^{-1})=0,
 \label{ektro}
\end{equation}
\begin{equation}
\pmb{Q}^{\sqrt{\frac{9}{4}-\lambda}}_{-\frac{1}{2}+\sqrt{M^2+4}}(-\alpha^{-1})\neq 0.
 \label{ektrqo}
\end{equation}
To end the proof of the characterization (\ref{transcendante}), we remark that the Wronskian formula (14.2.8) of \cite{nist} assures that (\ref{ektr}) implies (\ref{ektrq}) and  (\ref{ektro}) implies (\ref{ektrqo}), and we transform expression  (\ref{ektr}) using the recurrence relation ((14.10.5) in \cite{nist}),
$$
\left(P^{-\mu}_{\nu}(\cosh\rho)\right)'=-\frac{1}{\sinh^2\rho}\left((\nu-\mu)P^{-\mu}_{\nu-1}(\cosh\rho)-\nu\cosh\rho P^{-\mu}_{\nu}(\cosh\rho)\right).
$$
Now when $c=M=0$ the transcendent equation is simply $(3/2-\sqrt{9/4-\lambda})P_{-7/2}^{-\sqrt{9/4-\lambda}}(-\alpha^{-1})=0$.
We know that the function $x\mapsto P_{-7/2}^{-\sqrt{9/4-\lambda}}(-\alpha^{-1})(x)$ has no zero in $(1,\infty)$ when $\lambda\in[0,9/4[$ since $-7/2<-\sqrt{9/4-\lambda}<0$ (see 14.16 in \cite{nist}), thus we conclude that $0$ is the unique eigenvalue of $L_0$. Finally for $c=\infty$, $L_{\infty}$ is a non negative operator, hence we have to look for the eigenvalues in $[0,9/4[$. We recall that since  $-\sqrt{\frac{9}{4}-\lambda}\leq -\frac{1}{2}+\sqrt{M^2+4}$ for all $\lambda\in[0,9/4[$, $M\geq 0$, the equation $P^{-\sqrt{\frac{9}{4}-\lambda}}_{-\frac{1}{2}+\sqrt{M^2+4}}(x)= 0$ has no zero in $]1,\infty[$ if the integer part of $-\sqrt{\frac{9}{4}-\lambda}$ is even, and a unique zero when  the integer part of $-\sqrt{\frac{9}{4}-\lambda}$ is odd. That achieves the proof.

\fin

We now construct the spectral representation of $L_c$. If its point spectrum is not empty, we introduce the normalized eigenfunction $w(\rho;\lambda_j)\in H$ associated with a eigenvalue $\lambda_j\in\sigma_p(L_c)$, satisfying $Lw(.;\lambda_j)=\lambda_jw(.;\lambda_j)$, $\partial_{\rho}w(\rho_0)=cw(\rho_0)$,

\begin{equation}
w(\rho;\lambda_j):=\left\Vert P^{-\sqrt{\frac{9}{4}-\lambda_j}}_{-\frac{1}{2}+\sqrt{M^2+4}}\right\Vert^{-1}_{L^2\left(J',\frac{dx}{x^2-1}\right)} (\sinh\rho)^{-\frac{3}{2}} P^{-\sqrt{\frac{9}{4}-\lambda_j}}_{-\frac{1}{2}+\sqrt{M^2+4}}(\cosh\rho).
 \label{ujiro}
\end{equation}
For all $m>\frac{3}{2}$ we introduce the following generalized eigenfunction, solution of $Lw=m^2 w$, $\partial_{\rho}w(\rho_0)=cw(\rho_0)$ when $c\in\RR$ :
\begin{equation}
\begin{split}
w(\rho;m^2):=
&\frac{\sqrt{2m}}{\pi}(4m^2-9)^{\frac{1}{4}}\Gamma\left(\frac{1}{2}+\sqrt{M^2+4}-i\sqrt{m^2-\frac{9}{4}}\right) \left(\sinh\rho\right)^{-\frac{3}{2}}\\
&\left\vert\left(P_{-\frac{1}{2}+\sqrt{M^2+4}}^{i\sqrt{m^2-\frac{9}{4}}}\right)'(-\alpha^{-1})-\frac{\alpha}{\sqrt{1-\alpha^2}}\left(c-\frac{3}{2\sqrt{1-\alpha^2}}\right)P_{-\frac{1}{2}+\sqrt{M^2+4}}^{i\sqrt{m^2-\frac{9}{4}}}(-\alpha^{-1})\right\vert^{-1}\\
&\left\{\left[\left(P_{-\frac{1}{2}+\sqrt{M^2+4}}^{i\sqrt{m^2-\frac{9}{4}}}\right)'(-\alpha^{-1})-\frac{\alpha}{\sqrt{1-\alpha^2}}\left(c-\frac{3}{2\sqrt{1-\alpha^2}}\right)P_{-\frac{1}{2}+\sqrt{M^2+4}}^{i\sqrt{m^2-\frac{9}{4}}}(-\alpha^{-1})\right] {\pmb{Q}}_{-\frac{1}{2}+\sqrt{M^2+4}}^{-i\sqrt{m^2-\frac{9}{4}}}(\cosh\rho)\right.\\
-&\left.\left[\left({\pmb{Q}}_{-\frac{1}{2}+\sqrt{M^2+4}}^{-i\sqrt{m^2-\frac{9}{4}}}\right)'(-\alpha^{-1})-\frac{\alpha}{\sqrt{1-\alpha^2}}\left(c-\frac{3}{2\sqrt{1-\alpha^2}}\right){\pmb{Q}}_{-\frac{1}{2}+\sqrt{M^2+4}}^{-i\sqrt{m^2-\frac{9}{4}}}(-\alpha^{-1})\right] P_{-\frac{1}{2}+\sqrt{M^2+4}}^{i\sqrt{m^2-\frac{9}{4}}}(\cosh\rho)\right\},
\end{split}
 \label{ororw}
\end{equation}
and when $c=\infty$ :
\begin{equation}
\begin{split}
w(\rho;m^2):=
&\frac{\sqrt{2m}}{\pi}\left(4m^2-9\right)^{\frac{1}{4}}\Gamma\left(\frac{1}{2}+\sqrt{M^2+4}-i\sqrt{m^2-\frac{9}{4}}\right) \left(\sinh\rho\right)^{-\frac{3}{2}}\left\vert P_{-\frac{1}{2}+\sqrt{M^2+4}}^{i\sqrt{m^2-\frac{9}{4}}}(-\alpha^{-1})\right\vert^{-1}\\
&\left\{P_{-\frac{1}{2}+\sqrt{M^2+4}}^{i\sqrt{m^2-\frac{9}{4}}}(-\alpha^{-1}) {\pmb{Q}}_{-\frac{1}{2}+\sqrt{M^2+4}}^{-i\sqrt{m^2-\frac{9}{4}}}(\cosh\rho)-{\pmb{Q}}_{-\frac{1}{2}+\sqrt{M^2+4}}^{-i\sqrt{m^2-\frac{9}{4}}}(-\alpha^{-1}) P_{-\frac{1}{2}+\sqrt{M^2+4}}^{i\sqrt{m^2-\frac{9}{4}}}(\cosh\rho)\right\}.
\end{split}
 \label{ororwo}
\end{equation}

\begin{Proposition}
 \label{spectrepre}
For any $u\in H$, we put
\begin{equation}
C_j(u):=\left<u;w(.;\lambda_j)\right>_H,\;\;\Pi_pu:=\sum_{\lambda_j\in\sigma_p(L_c)}C_j(u)w(.;\lambda_j),\;\;\Pi_{ac}u:=u-\Pi_pu.
 \label{cj}
\end{equation}
The functions $w(\rho;m)$ are well defined continuous fonctions on $]0,\rho_0]_{\rho}\times]\frac{3}{2},\infty[_m$ and the  following limit exist,
\begin{equation}
u\in H,\;\;S\left(\Pi_{ac}u\right)(m):=\lim_{\epsilon\rightarrow0^+}\int_{\epsilon}^{\rho_0}\Pi_{ac}u(\rho)w(\rho;m^2)\sinh^2(\rho)\,d\rho\;\;in\;\;L^2\left(\frac{3}{2},\infty\right),
 \label{us}
\end{equation}
\begin{equation}
\hat{u}\in L^2\left(\frac{3}{2},\infty\right),\;\;\overline{S}\hat{u}(\rho):=\lim_{R\rightarrow\infty}\int_{\frac{3}{2}}^{R}\hat{u}(m)w(\rho;m^2)dm\;\;in\;\;H,
 \label{hatu}
\end{equation}
and we have :
\begin{equation}
u\in H,\;\;\overline{S}S\Pi_{ac}u=\Pi_{ac}u.
 \label{soso}
\end{equation}
Moreover, the map $u\longmapsto\left( (C_j(u))_j,S\left(\Pi_{ac}u\right)\right)$, is an isomorphism from $H$ onto $\CC^{\mid\sigma_p(L_c)\mid}\times L^2\left(\frac{3}{2},\infty\right)$ satisfying
\begin{equation}
u\in D_c\Rightarrow C_j(L_cu)=\lambda_jC_j(u),\;\;S\left(\Pi_{ac}L_cu\right)=m^2 S\left(\Pi_{ac}u\right),
 \label{domhat}
\end{equation}
\begin{equation}
u(\rho)=\sum_{\lambda_j\in\sigma_p(L_c)}C_j(u)w(\rho;\lambda_j)+\lim_{R\rightarrow\infty}\int_{\frac{3}{2}}^{R}S\left(\Pi_{ac}u\right) (m)w(\rho;m^2)dm\;\;in\;\;H,
 \label{decomp}
\end{equation}
\begin{equation}
\int_0^{\rho_0}\mid u(\rho)\mid^2\sinh^2(\rho)d\rho=\sum_{\lambda_j\in\sigma_p(L_c)}\mid C_j(u)\mid^2+\int_{\frac{3}{2}}^{\infty}\mid S\left(\Pi_{ac}u\right) (m)\mid^2dm.
 \label{isomo}
\end{equation}
 
\end{Proposition}

{\it Proof.} All the statements are well known. We only have to construct the spectral representation of $\tilde{L}_c$ with the analytic form involving the special functions. We use the following way (see \cite{pearson}, Theorem 7.4). We omit the details of the long computations, tedious but elementary. Given $\lambda>\frac{9}{4}$, we first determine a normalized upper solution $\varphi(\lambda;y)$ of $\tilde{L}\varphi=\lambda\varphi$ that is defined by
$$
\varphi(\lambda;y)=lim_{\epsilon\rightarrow 0^+}\varphi(\lambda+i\epsilon;y)\;\;in\;\;L^2_{loc}(y_0,\infty),
$$
where
$$
\forall\epsilon>0,\;\;\varphi(\lambda+i\epsilon;.)\in L^2(y_0,\infty),\;\;\tilde{L}\varphi(\lambda+i\epsilon,.)=(\lambda+i\epsilon) \varphi(\lambda+i\epsilon;.).
$$
From the asymptotic behaviours of the Legendre functions, we find
$$
\varphi(\lambda;y)=AP_{-\frac{1}{2}+\sqrt{M^2+4}}^{i\sqrt{\lambda-\frac{9}{4}}}(\coth y),\;\;A\in\CC.
$$
Moreover $\varphi(\lambda;.)$ has to satisfy
\begin{equation}
\varphi(\lambda;y)\overline{\partial_y\varphi(\lambda;y)}-\overline{\varphi(\lambda;y)}\partial_y\varphi(\lambda;y)=-i.
 \label{wronsfi}
\end{equation}
Since this wronskian does not depend on $y$, we make $y\rightarrow\infty$ and the asymptotics of the Legendre functions at $1^+$ allow to determine the constant of normalization $A$. We get
$$
A=(4\lambda-9)^{-\frac{1}{4}}.
$$
We note that (\ref{wronsfi}) implies that the real part and the imaginay part of $\varphi$ are linearly independent. As a consequence these real solutions cannot satisfy together the boundary condition at $y_0=0$. We deduce that for all $m>\frac{3}{2}$, $c\in\RR$, we have 
$$
P^{i\sqrt{m^2-\frac{9}{4}}}_{-\frac{1}{2}+\sqrt{M^2+4}}\left(-\frac{1}{\alpha}\right)\neq 0,\;\;
{\left(P^{i\sqrt{m^2-\frac{9}{4}}}_{-\frac{1}{2}+\sqrt{M^2+4}}\right)' \left(-\frac{1}{\alpha}\right)+\left(\frac{3}{2\sqrt{1-\alpha^2}}-c\right)\frac{\alpha}{\sqrt{1-\alpha^2}}P^{i\sqrt{m^2-\frac{9}{4}}}_{-\frac{1}{2}+\sqrt{M^2+4}}\left(-\frac{1}{\alpha}\right)}\neq 0,
$$
hence $w(\rho;m^2)$ is well defined as a continuous function on $]3/2,\infty[_m\times]0,\rho_0[_{\rho}$.

Now we compute the real valued solution $\tilde{v}(\lambda;y)$ of $\tilde{L}\tilde{v}=\lambda \tilde{v}$ satisfying the boundary condition $\tilde{v}'(\lambda;y_0)=\left(\frac{3}{2\alpha}-\frac{c \sqrt{1-\alpha^2}}{\alpha}\right)\tilde{v}(\lambda;y_0)$ and normalized to have the spectral amplitude $2\left\vert \tilde{v}\partial_y\varphi-\varphi\partial_y\tilde{v}\right\vert=1$. Since $\tilde{v}(\lambda;y)=v(\lambda;x)$, we use expression (\ref{ororv}) with $x^*=-\alpha^{-1}$ to get $\tilde{v}$ when $c\in\RR$ :
\begin{equation*}
\begin{split}
v(\lambda;x)&=\frac{1}{2}(4\lambda-9)^{\frac{1}{4}}\Gamma\left(\frac{1}{2}+\sqrt{M^2+4}-i\sqrt{\lambda-\frac{9}{4}}\right)\\
&\left\vert\left(P_{-\frac{1}{2}+\sqrt{M^2+4}}^{i\sqrt{\lambda-\frac{9}{4}}}\right)'(-\alpha^{-1})-\frac{\alpha}{\sqrt{1-\alpha^2}}\left(c-\frac{3}{2\sqrt{1-\alpha^2}}\right)P_{-\frac{1}{2}+\sqrt{M^2+4}}^{i\sqrt{\lambda-\frac{9}{4}}}(-\alpha^{-1})\right\vert^{-1}\\
&\left\{\left[\left(P_{-\frac{1}{2}+\sqrt{M^2+4}}^{i\sqrt{\lambda-\frac{9}{4}}}\right)'(-\alpha^{-1})-\frac{\alpha}{\sqrt{1-\alpha^2}}\left(c-\frac{3}{2\sqrt{1-\alpha^2}}\right)P_{-\frac{1}{2}+\sqrt{M^2+4}}^{i\sqrt{\lambda-\frac{9}{4}}}(-\alpha^{-1})\right] {\pmb{Q}}_{-\frac{1}{2}+\sqrt{M^2+4}}^{-i\sqrt{\lambda-\frac{9}{4}}}(x)\right.\\
&-\left.\left[\left({\pmb{Q}}_{-\frac{1}{2}+\sqrt{M^2+4}}^{-i\sqrt{\lambda-\frac{9}{4}}}\right)'(-\alpha^{-1})-\frac{\alpha}{\sqrt{1-\alpha^2}}\left(c-\frac{3}{2\sqrt{1-\alpha^2}}\right){\pmb{Q}}_{-\frac{1}{2}+\sqrt{M^2+4}}^{-i\sqrt{\lambda-\frac{9}{4}}}(-\alpha^{-1})\right] P_{-\frac{1}{2}+\sqrt{M^2+4}}^{i\sqrt{\lambda-\frac{9}{4}}}(x)\right\},
\end{split}
 \label{}
\end{equation*}
and when $c=\infty$ :
\begin{equation*}
\begin{split}
v(\lambda;x)=&\frac{1}{2}(4\lambda-9)^{\frac{1}{4}}\Gamma\left(\frac{1}{2}+\sqrt{M^2+4}-i\sqrt{\lambda-\frac{9}{4}}\right)\left\vert P_{-\frac{1}{2}+\sqrt{M^2+4}}^{i\sqrt{\lambda-\frac{9}{4}}}(-\alpha^{-1})\right\vert^{-1}\\
&\left\{P_{-\frac{1}{2}+\sqrt{M^2+4}}^{i\sqrt{\lambda-\frac{9}{4}}}(-\alpha^{-1}) {\pmb{Q}}_{-\frac{1}{2}+\sqrt{M^2+4}}^{-i\sqrt{\lambda-\frac{9}{4}}}(x)-{\pmb{Q}}_{-\frac{1}{2}+\sqrt{M^2+4}}^{-i\sqrt{\lambda-\frac{9}{4}}}(-\alpha^{-1}) P_{-\frac{1}{2}+\sqrt{M^2+4}}^{i\sqrt{\lambda-\frac{9}{4}}}(x)\right\}.
\end{split}
 \label{}
\end{equation*}

If we define for $\tilde{f}\in L^2(y_0,\infty)$, $C_j(\tilde{f}):=<\tilde{f};\tilde{u_j}>_{L^2}$, $\Pi_{ac}\tilde{f}:=f-\sum_{\lambda_j\in\sigma(L_c)}C_j(\tilde{f})\tilde{u}_j$ with $\tilde{u}_j(y):=w(\rho;\lambda_j)$ given by (\ref{ujiro}), the spectral theorem assures that the following limit exists,
$$
S\left(\Pi_{ac}\tilde{f}\right)(m):=\sqrt{\frac{2}{\pi}}\lim_{R\rightarrow\infty}\int_{y_0}^{R}\Pi_{ac}\tilde{f}(y)\tilde{V}(m;y)dy\;in\;L^2\left(\frac{3}{2},\infty\right),\;\tilde{V}(m;y):=2\sqrt{\frac{m}{\pi}}\tilde{v}(m^2;y),
$$
and we have
$$
\Pi_{ac}\tilde{f}(y)=\sqrt{\frac{2}{\pi}}\lim_{R\rightarrow\infty}\int_{\frac{3}{2}}^{R}S\left(\Pi_{ac}\tilde{f}\right)(m)\tilde{V}(m;y)dm\;in\;L^2(y_0,\infty),
$$
and the map $\tilde{f}\longmapsto\left( (C_j)_j,S\left(\Pi_{ac}\tilde{f}\right)\right)$, is an isometry from $L^2(y_0,\infty)$ onto $\CC^{\mid\sigma_p(L_c)\mid}\times L^2\left(\frac{3}{2},\infty\right)$.

Finally we put $w(\rho;m^2):=\sqrt{2/\pi}\left(\sinh\rho\right)^{-\frac{3}{2}}\tilde{V}\left(m;y=\log\left(\frac{e^{\rho}+1}{e^{\rho}-1}\right)\right)$, and we obtain formulas (\ref{ororw}),   (\ref{ororwo}) and the spectral representation of $L_c$.
\fin

We deduce a useful estimate of the norm of the Hilbert space $H_1$ defined by (\ref{hun}). 
\begin{Corollary}
If $u\in H_1$ the limit (\ref{decomp}) holds in $H_1$, moreover
 there exists $A,C>0$, independent of $u$, such that
\begin{equation}
C^{-1}\Vert u\Vert^2_{H_1}\leq \sum_{\lambda_j\in\sigma_p(L_c)}(A+\lambda_j)\mid C_j(u)\mid^2+\int_{\frac{3}{2}}^{\infty}(A+m^2)\mid S\left(\Pi_{ac}u\right) (m)\mid^2dm\leq C \Vert u\Vert^2_{H_1}.
 \label{ekivhm}
\end{equation}
 \label{ekivm}
\end{Corollary}

{\it Proof.} It is sufficient to establish (\ref{ekivhm}) for $u\in D_c$, and this formula follows from (\ref{ekivh}) and (\ref{isomo}). Then the convergence  of (\ref{decomp}) in $H_1$ is a straight consequence since
$$
\left\Vert u-\int_{\frac{3}{2}}^R S\left(\Pi_{ac}u\right) (m)w(\rho;m^2)dm\right\Vert_{H_1}^2\leq C \int_R^{\infty}(A+m^2)\mid S\left(\Pi_{ac}u\right) (m)\mid^2dm\underset{R\rightarrow\infty}{\longrightarrow} 0.
$$
\fin

The main result of this section is the following theorem that states the existence of Kaluza-Klein towers :

\begin{Theorem}
 \label{kktheo}
For any $c\in\RR\cup\{\infty\}$, the
 {\it finite energy solutions} $u$ of (\ref{eqq}) and (\ref{condlim}) can be expressed as
\begin{equation}
u(\tau,\mathbf{x},\rho)=\sum_{\lambda_j\in\sigma_p(L_c)}u_{\lambda_j}(\tau,\mathbf{x})w(\rho;\lambda_j)+\lim_{R\rightarrow\infty}\int_{\frac{3}{2}}^Ru_{m^2}(\tau,\mathbf{x})w(\rho;m^2)dm\;\;in\;\; C^0\left (\RR_{\tau}; X^1\right),
 \label{decomx}
\end{equation}
\begin{equation}
\partial_{\tau}u(\tau,\mathbf{x},\rho)=\sum_{\lambda_j\in\sigma_p(L_c)}\partial_{\tau}u_{\lambda_j}(\tau,\mathbf{x})w(\rho;\lambda_j)+\lim_{R\rightarrow\infty}\int_{\frac{3}{2}}^R\partial_{\tau}u_{m^2}(\tau,\mathbf{x})w(\rho;m^2)dm\;\;in\;\; C^0\left (\RR_{\tau}; X^0\right),
 \label{decomxo}
\end{equation}
where $w$ is defined by (\ref{ororw}), (\ref{ororwo}), $u_{\lambda_j}$ and $u_{m^2}$ are given by (\ref{ulambajumdeu}) below, and
\begin{equation}
u_{\lambda_j}\in C^0\left(\RR_{\tau};H^1\left(\RR^3_{\mathbf x}\right)\right)\cap C^1\left(\RR_{\tau};L^2\left(\RR^3_{\mathbf x}\right)\right),
 \label{reguj}
\end{equation}
and for any $T>0$
\begin{equation}
u_{m^2}\in L^2\left(\left]\frac{3}{2},\infty\right[_m;C^0\left([-T,T]_{\tau};H^1\left(\RR^3_{\mathbf{x}}\right)\right)\cap C^1\left([-T,T]_{\tau};L^2\left(\RR^3_{\mathbf{x}}\right)\right)\right),
 \label{estomak}
\end{equation}
\begin{equation}
mu_{m^2}\in L^2\left(\left]\frac{3}{2},\infty\right[_m;C^0\left([-T,T]_{\tau};L^2\left(\RR^3_{\mathbf{x}}\right)\right)\right).
 \label{estomakk}
\end{equation}
Moreover, $u_{\lambda_j}(\tau,\mathbf{x})w(\rho;\lambda_j)$ and for almost all $m$, $u_{m^2}(\tau,\mathbf{x})w(\rho;m^2)$ are solutions of (\ref{eqq}) and (\ref{condlim}), and $u_{\kappa}$ is solution of the Klein-Gordon equation (\ref{kgdsu}) in the Steady State space-time $dS_{\frac{1}{2}}^4$.

We have
\begin{equation}
\int_{\frac{3}{2}}^{\infty}\int_{\RR^3}\mid\partial_{\tau}u_{m^2}(\tau,\mathbf{x})\mid^2+\mid\nabla_{\mathbf x}u_{m^2}(\tau,\mathbf{x})\mid^2+m^2\mid u_{m^2}(\tau,\mathbf{x})\mid^2d\mathbf{x}dm\lesssim \Vert u(\tau)\Vert_{X^1}^2+\Vert\partial_{\tau}u(\tau)\Vert_{X^0}^2,
 \label{naurme}
\end{equation}
and the energy (\ref{ener}) of $u$ is the sum of the energy (\ref{energstead}) of modes $u_{\kappa}$ :
\begin{equation}
\mathcal{E}(u,\tau)=\sum_{\lambda_j\in\sigma_p(L_c)}\mathcal{E}_{\lambda_j}(u_{\lambda_j},\tau)+\int_{\frac{3}{2}}^{\infty}\mathcal{E}_{m^2}(u_{m^2},\tau)dm.
 \label{somener}
\end{equation}
\end{Theorem}


{\it Proof.} 
Given a finite energy solution $u$, for all $\tau\in\RR$, and almost all ${\mathbf x}\in\RR^3$, the map $\rho\mapsto u(\tau,\mathbf{x}, \rho)$ belongs to $H_1$, and the map  $\rho\mapsto \partial_{\tau}u(\tau,\mathbf{x}, \rho)$ belongs to $H$. Hence with the notations of (\ref{cj}), (\ref{us}), (\ref{hatu}), we introduce
\begin{equation}
u_{\lambda_j}(\tau,\mathbf{x}):=C_j(u(\tau,\mathbf{x},.)),\;\;u_{m^2}(\tau,\mathbf{x}):=S\left(\Pi_{ac}u(\tau,\mathbf{x},.)\right)(m).
 \label{ulambajumdeu}
\end{equation}
It is clear by (\ref{isomo}) and (\ref{ekivhm}) that $u\mapsto (u_{\lambda_j}, u_{m^2})$ is continuous from $C^0\left(\RR_{\tau};X^1\right)\cap C^1\left(\RR_{\tau};X^0\right)$ to $C^0(\RR_{\tau};H^1(\RR^3_{\mathbf{x}}))\cap C^1(\RR_{\tau};L^2(\RR^3_{\mathbf x}))\times\Theta$ where
$$
\Theta:=\left\{v;\;v,\partial_{\tau}v,\;\nabla_{\mathbf{x}}v,\;mv\in  C^0\left(\RR_{\tau};L^2\left(]3/2,\infty[_m\times\RR^3_{\mathbf{x}}\right)\right)\right\}.
$$
Hence Corollary \ref{ekivm} and (\ref{decomp}) assure that for all fixed $\tau$ the limits (\ref{decomx}) and (\ref{decomxo}) hold respectively in $X^1$ and in $X^0$.
Moreover (\ref{isomo}) and (\ref{ekivhm}) imply that
\begin{equation*}
\begin{split}
\Vert\int_{\frac{3}{2}}^M(u_{m^2}(\tau)-u_{m^2}(\sigma))w(m^2)dm\Vert_{X^1}+
\Vert\int_{\frac{3}{2}}^M(&\partial_{\tau}u_{m^2}(\tau)-\partial_{\tau}u_{m^2}(\sigma))w(m^2)dm\Vert_{X^0}\\
&\leq C \left( \Vert u(\tau)-u(\sigma)\Vert_{X^1}+\Vert \partial_{\tau}u(\tau)-\partial_{\tau}u(\sigma)\Vert_{X^0}\right)
\end{split}
\end{equation*}
therefore $\int_{\frac{3}{2}}^Ru_{m^2}(\tau,\mathbf{x})w(\rho;m^2)dm$ and $\int_{\frac{3}{2}}^R\partial_{\tau}u_{m^2}(\tau,\mathbf{x})w(\rho;m^2)dm$ are equicontinuous families of functions in $C^0\left (\RR_{\tau}; X^1\right)$ and  $C^0\left (\RR_{\tau}; X^0\right)$ respectively.
We conclude that they are converging in these spaces.

The estimate (\ref{naurme}) follows from (\ref{isomo}) and (\ref{ekivhm}). Now it is sufficient to prove (\ref{somener}) for the dense subset of the strong solutions that are in $C^0(\RR_{\tau};X^2_c)$. Since $X^2_c\subset L^2(\RR^3_{\mathbf x};D_c)$ we can write
$$
{\mathcal E}(u,\tau)=\int_{\RR^3}\Vert\partial_{\tau} u(\tau,\mathbf{x},.)\Vert^2_H+e^{-2\tau}\Vert \nabla_{\mathbf x}u(\tau,\mathbf{x},.)\Vert^2_H+\left<L_cu(\tau,\mathbf{x},.);u(\tau,\mathbf{x},.)\right>_Hd\mathbf{x}
$$
and (\ref{somener})  follows from (\ref{domhat}) and (\ref{isomo}). By the same argument of density, it is sufficient to prove that $u_{\lambda_j}$ is solution of the Klein-Gordon equation in the steady state universes in the case of a strong solution $u$. We take a test function $\Phi\in C^{\infty}_0(\RR_{\tau}\times\RR^3_{\mathbf x})$ and we choose a sequence $w_n(\rho;\lambda_j)\in C^{\infty}_0(]0,\rho_0[)$ that tends to $w(\rho;\lambda_j)$ in $H$ as $n\rightarrow\infty$. Denoting $<;>_{x}$ the bracket of distributions with respect to the $x$-variable, we write
\begin{equation*}
\begin{split}
\left<(\partial^2_{\tau}+3\partial_{\tau}-e^{-2\tau}\Delta_{\mathbf x})u_{\lambda_j};\Phi\right>_{\tau,\mathbf{x}}&=
\left<u_{\lambda_j};(\partial^2_{\tau}-3\partial_{\tau}-e^{-2\tau}\Delta_{\mathbf x})\Phi\right>_{\tau,\mathbf{x}} \\
&=\int u(\tau,\mathbf{x},\rho)w(\rho;\lambda_j) (\partial^2_{\tau}-3\partial_{\tau}-e^{-2\tau}\Delta_{\mathbf x})\Phi(\tau,\mathbf{x})\sinh^2(\rho)d\tau d\mathbf{x}d\rho\\
&=\lim_n\left<(\partial^2_{\tau}+3\partial_{\tau}-e^{-2\tau}\Delta_{\mathbf x})u;\Phi\otimes w_n\right>_{\tau,\mathbf{x},\rho}\\
&=\lim_n\left<-Lu;\Phi\otimes w_n\right>_{\tau,\mathbf{x},\rho}\\
&=\int\left<-L_cu(\tau,\mathbf{x},.);w(.;\lambda_j)\right>_H\Phi(\tau,\mathbf{x})d\tau d\mathbf{x}\\
&=-\lambda_j\left<u_{\lambda_j}; \Phi\right>_{\tau,\mathbf{x}}
\end{split}
\end{equation*}
hence we deduce that $(r\partial^2_{\tau}+3\partial_{\tau}-e^{-2\tau}\Delta_{\mathbf x}+\lambda_j)u_{\lambda_j}=0$. Now we prove that $u_{m^2}$ is solution of the equation
\begin{equation}
(\partial^2_{\tau}+3\partial_{\tau}-e^{-2\tau}\Delta_{\mathbf x}+m^2)\varphi=0\;\;in\;\;{\mathcal D}'(\RR_{\tau}\times\RR^3_{\mathbf x}\times]3/2,\infty[_m).
 \label{ekrem}
\end{equation}
We take $\Psi\in C^{\infty}_0(\RR_{\tau}\times\RR^3_{\mathbf x}\times]3/2,\infty[_m)$, and we compute
\begin{equation*}
\begin{split}
\left<(\partial^2_{\tau}+3\partial_{\tau}-e^{-2\tau}\Delta_{\mathbf x})u_{m^2};\Psi\right>_{\tau,\mathbf{x},m}&=
\left<u_{m^2};(\partial^2_{\tau}-3\partial_{\tau}-e^{-2\tau}\Delta_{\mathbf x})\Psi\right>_{\tau,\mathbf{x},m} \\
&=\int \left[\int_0^{\rho_0}\Pi_{ac}u(\tau,\mathbf{x},\rho)(\partial^2_{\tau}-3\partial_{\tau}-e^{-2\tau}\Delta_{\mathbf x})\overline{S}\Psi(\tau,\mathbf{x},\rho)\sinh^2(\rho)d\rho\right]d\tau d\mathbf{x}\\
&=\int_0^{\rho_0}\left[\int \Pi_{ac}(\partial^2_{\tau}+3\partial_{\tau}-e^{-2\tau}\Delta_{\mathbf x})u(\tau,\mathbf{x},\rho) \overline{S}\Psi(\tau,\mathbf{x},\rho) d\tau d\mathbf{x}\right]d\rho\\
&=-\int\left[\int_0^{\rho_0}\Pi_{ac}u(\tau,\mathbf{x},\rho)L_c\overline{S}\Psi(\tau,\mathbf{x},\rho)\sinh^2(\rho)d\rho\right]d\tau d\mathbf{x}\\
&=-\int m^2u_{m^2}(\tau,\mathbf{x})\Psi(\tau,\mathbf{x},m)d\tau d\mathbf{x} dm,
\end{split}
\end{equation*}
thus (\ref{ekrem}) is established. Now the Cauchy problem for this equation is well posed in
$\Theta$. The uniqueness follows from the energy  estimate 
\begin{equation}
\int_{\frac{3}{2}}^{\infty}\mathcal{E}_{m^2}(\varphi(.,m),\tau)dm=\int_{\frac{3}{2}}^{\infty}\mathcal{E}_{m^2}(\varphi(.,m),\tau_*)dm-2\int_{\tau_*}^{\tau}\int_{\RR^3}\int_{\frac{3}{2}}^{\infty}3\mid\partial_{\tau}\varphi\mid^2+e^{-2\tau}\mid\nabla_{\mathbf x}\varphi\mid^2d\mathbf{x}dm d\sigma
 \label{estermam}
\end{equation}
that is easily get by integration by parts when
\begin{equation}
\left\{
\begin{split}
&\varphi\in C^1\left(\RR_{\tau};L^2\left(]3/2,\infty[_m;H^1(\RR^3_{\mathbf{x}})\right)\right)\cap C^2\left(\RR_{\tau};L^2\left(]3/2,\infty[_m;L^2(\RR^3_{\mathbf{x}})\right)\right),\\
&m\varphi\in C^1\left(\RR_{\tau};L^2\left(]3/2,\infty[_m;L^2(\RR^3_{\mathbf{x}})\right)\right).
\end{split}
\right.
 \label{galoupi}
\end{equation}
In the general case where $\varphi\in\Theta$, we introduce $\varphi_{\epsilon,M}(\tau,\mathbf{x},m))=\mathbf{1}_{[\frac{3}{2},M]}(m)\int\theta_{\epsilon}(\mathbf{x}-\mathbf{y})\varphi(\tau,\mathbf{y},m)d\mathbf{y}$ where $\theta_{\epsilon}$ is a mollifiers sequence in $\RR^3$. It is clear that  $\varphi_{\epsilon,M}(\tau,\mathbf{x},m))$ is a solution of (\ref{ekrem}) that tends to $\varphi$ in $\Theta$ as $\epsilon\rightarrow 0$, $M\rightarrow\infty$, and satisfies (\ref{galoupi}). Therefore we get (\ref{estermam}) and the Gronwall lemma shows that $\varphi=0$ if $\varphi(\tau_*,.)=\partial_{\tau}\varphi(\tau_*,.)=0$. Now given $\tau_*$, for all almost $m\geq 3/2$, we have $u_{m^2}(\tau_*,.)\in H^1(\RR^3)$, $\partial_{\tau}u_{m^2}(\tau_*,.)\in L^2(\RR^3)$, hence we can apply the Theorem \ref{teosteady} and we get $v_{m^2}\in C^0(\RR_{\tau};H^1(\RR^3))\cap C^1(\RR_{\tau};L^2(\RR^3))$ solution of  (\ref{kgdsu}), (\ref{initialdsu}) with $\kappa=m^2$, $v_{m^2}(\tau_*)=u_{m^2}(\tau_*)$, $\partial_{\tau}v_{m^2}(\tau_*)=\partial_{\tau}u_{m^2}(\tau_*)$. The energy estimate (\ref{energstead}) implies that
$$
mv_{m^2}\in L^2\left(]3/2,\infty[_m;C^0(\RR_{\tau};L^2(\RR^3))\right),
$$
$$
v_{m^2}\in L^2\left(]3/2,\infty[_m;C^0(\RR_{\tau};H^1(\RR^3))\cap C^1(\RR_{\tau};L^2(\RR^3))\right)
$$
and since this space is included in $\Theta$, the uniqueness assures that $u_{m^2}=v_{m^2}$, hence (\ref{estomak}) and (\ref{estomakk}) are proved.
\fin


\section{Gravitational Waves}

In this section we consider the very important case of the gravitational waves that are described by equation (\ref{eq}) with $M=0$ and the boundary condition on the De Sitter brane is of Neumann type, {\it i.e.} (\ref{cl}) with $c=0$.
The crucial point is the existence of the sector of the massless graviton that is just the set of the solutions
$$
u_{\varphi}(t,\mathbf{x},z):=\varphi\left(-\frac{1}{2}\log\left(t^2-z^2\right),\mathbf{x}\right),
$$
where $\varphi=\varphi(\tau,{\mathbf x})$ is any solution of the massless wave equation in the Steady State space-time $dS^4_{\frac{1}{2}}$,
$$
\left(\partial_{\tau}^2+3\partial_{\tau}-e^{-2\tau}\Delta_{\mathbf x}\right)\varphi=0,\;\;\tau\in\RR,\;\;{\mathbf x}\in\RR^3.
$$
The graviton $u_{\varphi}$ is solution of the massless wave equation on $AdS^5$, $\left(\partial_t^2-\Delta_{\mathbf x}-\partial_z^2+\frac{3}{z}\partial_z\right)u_{\varphi}=0$ in $\mathcal{O}=\Omega\times\RR^3_{\mathbf x}$,  $\Omega:=\left\{(t,z); t<-z<0\right\}$, and satisfies the Neumann condition $\alpha\partial_tu_{\varphi}+\partial_zu_{\varphi}=0$ on $z=\alpha t$, $t<0$, for any $\alpha\in]-1,0[$. In the $\tau,\rho$ coordinates, it is just a solution independent of $\rho$, {\it i.e.} $u_{\varphi}(\tau,\mathbf{x},\rho)=\varphi(\tau,{\mathbf x})$ (by abuse of notation we write $u_{\varphi}(\tau,\mathbf{x},\rho):=u_{\varphi}(t,\mathbf{x},z)$). When $\varphi$ is a finite energy solution, {\it i.e.} $\varphi\in C^0\left(\RR_{\tau};H^1(\RR^3_{\mathbf x})\right)\cap C^1\left(\RR_{\tau};L^2(\RR^3_{\mathbf x})\right)$, then $u_{\varphi}\in C^0\left(\RR_{\tau};X^1\right)\cap C^1(\RR_{\tau};X^0)\cap\mathcal{D}'\left(\RR_{\tau}\times\RR^3_{\mathbf x};D_c\right)$. Moreover, we have $u_{\varphi}\in C^0\left(\Omega;H^1(\RR^3_{\mathbf x})\right)$, $\nabla_{t,\mathbf{x},z}u_{\varphi}\in C^0\left(\Omega;L^2(\RR^3_{\mathbf x})\right)$. It is interesting to investigate the behaviour of this graviton along the De Sitter brane $\mathcal{B}$  as $ t\rightarrow 0^-$ , and when we approach its Cauchy horizon $\mathcal{N}$. We introduce
$$
\gamma:=\sqrt{\frac{2}{\sinh\rho_0\cosh\rho_0-\rho_0}}.
$$

\begin{Proposition}
For any finite energy massless graviton $u_{\varphi}$, there exists $\phi\in H^2(\RR^3_{\mathbf x})$ such that when $e^{-2\tau}=t^2-z^2\rightarrow 0$ with
$t<-z< 0$, we have :
 \label{propufi}
 \begin{equation*}
\gamma\Vert u_{\varphi}(\tau,.)-\phi(.)\Vert_{X^1}=\Vert u_{\varphi}(t,.,z)-\phi\Vert_{H^1(\RR^3_{\mathbf{x}})}\rightarrow 0,
 \label{}
\end{equation*}
\begin{equation*}
\gamma\left\Vert e^{2\tau}\partial_{\tau}u_{\varphi}(\tau,.)-\Delta\phi\right\Vert_{X^0}=\left\Vert \frac{1}{t}\partial_tu_{\varphi}(t,.,z)+\Delta\phi\right\Vert_{L^{2}(\RR^3_{\mathbf{x}})}=\left\Vert \frac{1}{z}\partial_zu_{\varphi}(t,.,z)-\Delta\phi\right\Vert_{L^{2}(\RR^3_{\mathbf{x}})}\rightarrow 0,
 \label{}
\end{equation*}
\begin{equation*}
\left\Vert \frac{1}{z-t}\left(\partial_t+\partial_z\right)u_{\varphi}(t,.,z)-\Delta\phi\right\Vert_{L^{2}(\RR^3_{\mathbf{x}})}\rightarrow 0,
 \label{}
\end{equation*}
\begin{equation*}
\left\Vert \frac{1}{z+t}\left(\partial_t-\partial_z\right)u_{\varphi}(t,.,z)+\Delta\phi\right\Vert_{L^{2}(\RR^3_{\mathbf{x}})}\rightarrow 0,
 \label{}
\end{equation*}
and the Fourier transform of $\phi$ is given by
\begin{equation}
\hat{\phi}(\pmb\xi)=\sqrt{\frac{\pi}{2}}\mid\pmb\xi\mid^{-\frac{1}{2}}\left\{J_{\frac{1}{2}}(\mid\pmb\xi\mid)\hat{\varphi}(0,\pmb\xi)+J_{\frac{3}{2}}(\mid\pmb\xi\mid)\mid\pmb\xi\mid^{-1}\partial_{\tau}\hat{\varphi}(0,\pmb\xi)\right\}.
 \label{forfi}
\end{equation}
\end{Proposition}

{\it Proof.}
 It is a direct application of the theorem \ref{teosteady} and the  formula (\ref{pifi}). Since $u_{\varphi}(t,\mathbf{x},z)=u_{\varphi}(\tau,\mathbf{x},\rho)=\varphi(\tau,\mathbf{x})$ with $\tau=-\frac{1}{2}\ln(t^2-z^2)$, we have
$$
\frac{1}{z}\partial_zu_{\varphi}(t,\mathbf{x},z)=-\frac{1}{t}\partial_tu_{\varphi}(t,\mathbf{x},z)=e^{2\tau}\partial_{\tau}\varphi(\tau,\mathbf{x}),
$$
and the asymptotics follow from (\ref{assds}). The expression of $\hat{\phi}$ is given by  the  formula (\ref{pifi}) with $\tau_*=0$.
\fin

The main result of this part states that the usual De Sitter gravity is recovered on the brane : the leading term of any gravitational fluctuation in the $AdS$ bulk $\mathcal{M}$ is a massless graviton propagating on the brane. In this sense, we may conclude that a De Sitter brane in an Anti-de Sitter bulk is linearly stable.

\begin{Theorem}
 \label{theofluc}
We assume $M=0$, $c=0$. Then given $u_0\in X^1$, $u_1\in X^0$, there exists a solution $\varphi\in C^0\left(\RR_{\tau};H^1(\RR^3_{\mathbf x})\right)\cap C^1\left(\RR_{\tau};L^2(\RR^3_{\mathbf x})\right)$ of the massless wave equation (\ref{eqds}) in the Steady State Universe $dS^4_{\frac{1}{2}}$ such that the solution $u\in  C^1\left(\RR_{\tau};X^0\right)\cap C^0\left(\RR_{\tau};X^1\right)\cap{\mathcal D}'\left(\RR_{\tau}\times\RR^3_{\mathbf x};D_{c}\right)$ of (\ref{eqq}), (\ref{condlim}) and (\ref{condinit}), satisfies
\begin{equation}
\left\Vert (u-u_{\varphi})(\tau)\right\Vert_{X^1}+\left\Vert \partial_{\tau}u(\tau)\right\Vert_{X^0}\longrightarrow 0,\;\;\tau\rightarrow+\infty,
 \label{deka}
\end{equation}
\begin{equation}
\left\Vert \left(\partial_{\tau}+\frac{3}{2}\right)(u-u_{\varphi})(\tau)\right\Vert_{X^0}=O(e^{-\frac{3}{2}\tau}).
 \label{dekaa}
\end{equation}
The initial data of $\varphi$ are given by
\begin{equation}
\varphi(\tau_*,\mathbf{x})=\gamma^2\int_0^{\rho_0}u_0(\mathbf{x},\rho)\sinh^2\rho d\rho\in H^1(\RR_{\mathbf x}^3),\;\partial_{\tau}\varphi(\tau_*,\mathbf{x})=\gamma^2\int_0^{\rho_0}u_1(\mathbf{x},\rho)\sinh^2\rho d\rho\in L^2(\RR_{\mathbf x}^3).
 \label{condinitfi}
\end{equation}
\end{Theorem}

{\it Proof of Theorem \ref{theofluc}.}

By (\ref{foncprop}) and (\ref{pointspectrum}) we know that the point spectrum of $L_0$ is $\{0\}$ and a normalized eigenfunction is the constant function
$
w(\rho;0)=\gamma$.
Theorem \ref{kktheo} assures that
$$
u(\tau,\mathbf{x},\rho)=u_{\varphi}(\tau,\mathbf{x},\rho)
+\lim_{R\rightarrow\infty}\int_{\frac{3}{2}}^Ru_{m^2}(\tau,\mathbf{x})w(\rho;m^2)dm\;\;in\;\; C^0\left (\RR_{\tau}; X^1\right)\cap C^1\left(\RR_{\tau};X^0\right),
$$
where $u_{\varphi}(\tau,\mathbf{x},\rho)=\varphi(\tau,\mathbf{x})$ solution of (\ref{eqds}) with the initial data (\ref{condinitfi}).
We also have
$$
\left\Vert(u-u_{\varphi})(\tau,\mathbf{x},.)\right\Vert_H^2=\int_{\frac{3}{2}}^{\infty}\mid u_{m^2}(\tau,\mathbf{x})\mid^2dm,\;\;
\left\Vert\nabla_{\tau,\mathbf{x}}(u-u_{\varphi})(\tau,\mathbf{x},.)\right\Vert_H^2=\int_{\frac{3}{2}}^{\infty}\mid \nabla_{\tau,\mathbf{x}}u_{m^2}(\tau,\mathbf{x})\mid^2dm,
$$
and by (\ref{ekivhm}) we also have
$$
\left\Vert(u-u_{\varphi})(\tau,\mathbf{x},.)\right\Vert_{H_1}^2\leq C\int_{\frac{3}{2}}^{\infty}m^2\mid u_{m^2}(\tau,\mathbf{x})\mid^2dm.
$$
Then we deduce from the Fubini theorem that
$$
\left\Vert (u-u_{\varphi})(\tau)\right\Vert^2_{X^1}+\left\Vert \partial_{\tau}(u-u_{\varphi})(\tau)\right\Vert^2_{X^0}\leq 
C\int_{\frac{3}{2}}^{\infty}\Vert \partial_{\tau}u_{m^2}(\tau)\Vert_{L^2}^2+
\Vert \nabla_{\mathbf x}u_{m^2}(\tau)\Vert_{L^2}^2+m^2\Vert u_{m^2}(\tau)\Vert_{L^2}^2 dm.
$$
From the Theorem \ref{teosteady} we know that for any $m>\frac{3}{2}$, 
the integrand of this integral tends to zero as $\tau\rightarrow\infty$ and is bounded uniformly with respect to $\tau>\tau_*$ by 
$$
e^{3\mid\tau_*\mid}
\left(\Vert \partial_{\tau}u_{m^2}(\tau_*)\Vert_{L^2}^2+
\Vert \nabla_{\mathbf x}u_{m^2}(\tau_*)\Vert_{L^2}^2+m^2\Vert u_{m^2}(\tau_*)\Vert_{L^2}^2\right)
$$
which belongs to $L^1(3/2,\infty)$ by (\ref{naurme}).
We conclude by the dominated convergence theorem that  $\left\Vert (u-u_{\varphi})(\tau)\right\Vert_{X^1}+\left\Vert \partial_{\tau}(u-u_{\varphi})(\tau)\right\Vert_{X^0}$ tends to $ 0$ as $\tau\rightarrow+\infty$. Since  $\left\Vert \partial_{\tau}u_{\varphi}(\tau)\right\Vert_{X^0}= O(e^{-2\tau})$ by (\ref{assds}),  (\ref{deka}) is satisfied. Finally (\ref{dekaa}) is a direct consequence of (\ref{enert}) and  (\ref{naurme}).
\fin

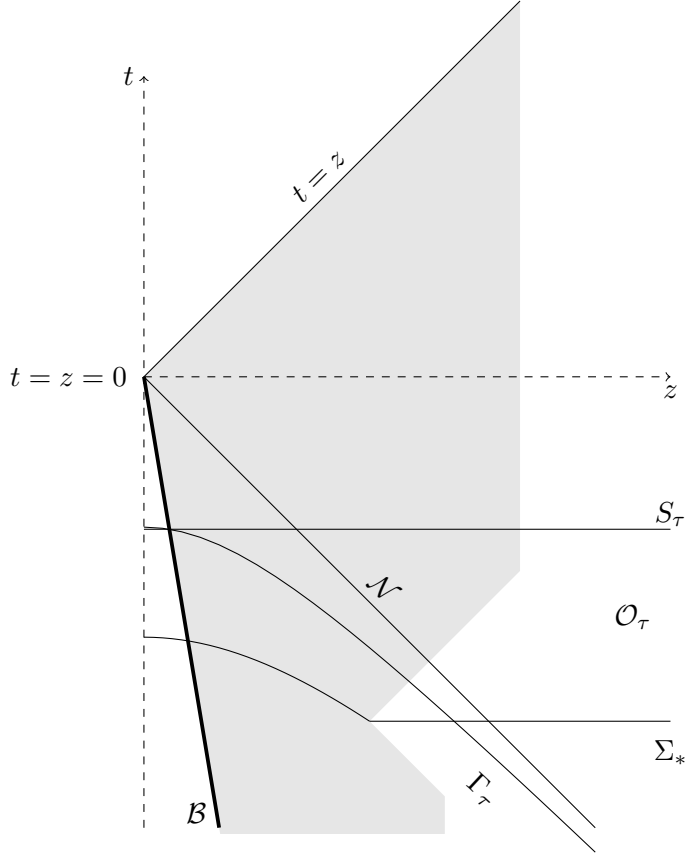
\begin{figure}
\begin{tikzpicture}
\draw  [->] [dashed] (0,-2) -- (0,8);
\fill[color=gray!20]
(0,4) -- (5,9)
-- (5,{6-sqrt(21)})--(3,{4-sqrt(21)})--(4,{3-sqrt(21)})--(4,{2.5-sqrt(21)})--({sqrt(21/36)+1.5/6},{2.5-sqrt(21)});
\draw [->]   [dashed] (0,4) -- (7,4);
\draw (7,3.8) node {$z$};
\draw (-0.2,8) node {$t$};
\draw (-1,4) node{$t=z=0$};
\draw (0,4) -- (5,9) node[midway,sloped,above] {$t=z$};
\draw (0,4) -- (6,-2) node[midway,sloped,above] {${\mathcal N}$};
\draw (7,-1) node {$\Sigma_*$};
\draw (6.5,0.8) node {${\mathcal O}_{\tau}$};
\draw [line width=0.5mm] [domain=0:1] plot (\x, {4-6*\x});
\draw [domain=0:6] plot (\x, {4-sqrt((\x)^2+4)});
\draw [domain=0:3] plot (\x, {4-sqrt((\x)^2+12)});
\draw (3,{4-sqrt(21)})--(7,{4-sqrt(21)});
\draw (0,{4-12/sqrt(35)})-- (7,{4-12/sqrt(35)});
\draw (7,{4.2-12/sqrt(35)}) node{$S_{\tau}$};
\draw (0.7,-1.8) node{$\mathcal B$};
\draw (4.5,-1.5) node[rotate=-43] {$\Gamma_{\tau}$};

\end{tikzpicture}
\caption{The solution is supported in the grey zone.}
\label{desdus}
\end{figure}


We end this work by investigating the gravitational fluctuations beyond the Cauchy horizon $\mathcal{N}$. We obviously have to add some constraint when $z>-t$. Roughly speaking, we suppose that there is no field incoming from the past null infinity, {\it i.e.} $u(t,\mathbf{x},z)\rightarrow 0$ as $t\rightarrow-\infty$, $z+t=Cst.>0$. More precisely we consider initial data $u_0,u_1\in C_0^{\infty}\left(\RR^3_{\mathbf x}\times (\epsilon,\rho_0)_{\rho}\right)$, $0<\epsilon<\rho_0$ and $u(\tau,\mathbf{x},\rho)$ the solution of  (\ref{eqq}), (\ref{condlim}) and (\ref{condinit}) in $\mathcal{M}$ with $M=c=0$. We define a Cauchy hypersurface $\Sigma_*$ in a domain $\tilde{\mathcal M}$ larger than  $\mathcal M $ (see Figure \ref{desdus}) :
\begin{equation*}
\tilde{\mathcal M}:=\left\{(t,\mathbf{x},z)\in\RR\times\RR^3\times]0,\infty[;\;\;\max(\alpha t,t)<z\right\},
 \label{}
\end{equation*}
\begin{equation*}
\Sigma_*:=\left(\{\tau_*\}\times\RR^3_{\mathbf x}\times[\epsilon,\rho_0[_{\rho}\right)\cup\left(\left\{t=-\frac{\cosh\epsilon}{\sinh\epsilon}e^{-\tau_*}\right\}\times\RR^3_{\mathbf x}\times[\frac{1}{\sinh\epsilon}e^{-\tau_*},\infty[_z\right),
 \label{}
\end{equation*}
It is clear that $\Sigma_*$ is a Cauchy hypersurface of $\tilde{\mathcal{M}}$ that is globally hyperbolic and its timelike boundary is the De Sitter brane $\mathcal{B}$. We define $\tilde{u}_j=u_j$ on $\{\tau_*\}\times\RR^3_{\mathbf x}\times[\epsilon,\rho_0[_{\rho}$ and $\tilde{u}_j=0$ on $\left\{t=-\frac{\cosh\epsilon}{\sinh\epsilon}e^{-\tau_*}\right\}\times\RR^3_{\mathbf x}\times[\frac{1}{\sinh\epsilon}e^{-\tau_*},\infty[_z$. By the standard results on the hyperbolic mixed problem (see {\it e.g.} \cite{ladyz}), there exists $\tilde{u}\in C^{\infty}(\tilde{\mathcal M}\cup\mathcal{B})$ solution of  $\left(\partial_t^2-\Delta_{\mathbf x}-\partial_z^2+\frac{3}{z}\partial_z\right)\tilde{u}=0$ in $\tilde{\mathcal{M}}$, satisfying the Neumann condition $\alpha\partial_t\tilde{u}+\partial_z\tilde{u}=0$ on $\mathcal{B}$, and the Cauchy condition $\tilde{u}=\tilde{u}_0$ on $\Sigma_*$,  $\partial_{\tau}\tilde{u}=\tilde{u}_1$ on $\{\tau_*\}\times\RR^3_{\mathbf x}\times[\epsilon,\rho_0[_{\rho}$ and $\partial_t\tilde{u}=0$ on  $\left\{t=-\frac{\cosh\epsilon}{\sinh\epsilon}e^{-\tau_*}\right\}\times\RR^3_{\mathbf x}\times[\frac{1}{\sinh\epsilon}e^{-\tau_*},\infty[_z$. $\tilde{u}$ is supported in $\left\{(t,\mathbf{x},z)\in\RR\times\RR^3\times]0,\infty[;\;\;\max(\alpha t,t)\leq z\leq \frac{1}{\sinh\epsilon}e^{-\tau_*}+\left\vert t+\frac{\cosh\epsilon}{\sinh\epsilon}e^{-\tau_*}\right\vert\right\}$ and we obviously have $\tilde{u}(t,\mathbf{x},z)=u(\tau,\mathbf{x},\rho)$ in $\mathcal{M}$. We study the energy on the hypersurface $t=Cst.$

\begin{Theorem}
 \label{bloww}
 We assume that $\phi$ defined by (\ref{forfi}) and (\ref{condinitfi}) is non zero. Then we have
\begin{equation*}
\int_{\alpha t}^{\infty}\int_{\RR^3}\mid\nabla_{t,{\mathbf x},z}\tilde{u}(t,\mathbf{x},z)\mid^2z^{-3}d\mathbf{x}dz\gtrsim \frac{1}{t^2},\;\;t\rightarrow 0^-.
 \label{}
\end{equation*}

\end{Theorem}

{\it Proof of Theorem \ref{bloww}.} Since $\left(\partial_t^2-\Delta_{\mathbf x}-\partial_z^2+\frac{3}{z}\partial_z\right)\tilde{u}=0$ in $\tilde{\mathcal{M}}$, we have $$\nabla_{t,{\mathbf x},z}.\left(z^{-3}\mid\nabla_{t,{\mathbf x},z}\tilde{u}\mid^2,-2z^{-3}\partial_t\tilde{u}\nabla_{{\mathbf x},z}\tilde{u}\right)=0\;\; in \;\;\tilde{\mathcal{M}}.$$We integrate this divergence in the domain
\begin{equation*}
{\mathcal O}_{\tau}:=\left\{(t,\mathbf{x},z)\in\RR^-\times\RR^3\times\left[-\frac{\alpha}{\sqrt{1-\alpha^2}}e^{-\tau},\infty\right[;\;-\sqrt{e^{-2\tau}+z^2}<t<-\frac{e^{-\tau}}{\sqrt{1-\alpha^2}}\right\}.
 \label{}
\end{equation*}
of which the boundary is $\Gamma_{\tau}\cup S_{\tau}$ defined by :
\begin{equation*}
\Gamma_{\tau}:=\{\tau\}\times\RR^3_{\mathbf x}\times]0,\rho_0[=\left\{(t,\mathbf{x},z)\in\RR^-\times\RR^3_{\mathbf x}\times\left[-\frac{\alpha}{\sqrt{1-\alpha^2}}e^{-\tau},\infty\right[;\;t^2-z^2=e^{-2\tau}\right\},
 \label{}
\end{equation*}
\begin{equation*}
S_{\tau}:=\left\{t=-\frac{e^{-\tau}}{\sqrt{1-\alpha^2}}\right\}\times\RR^3_{\mathbf x}\times\left[-\frac{\alpha}{\sqrt{1-\alpha^2}}e^{-\tau},\infty\right[_z.
 \label{}
\end{equation*}
We get by the Ostrogradski theorem :
\begin{equation*}
\begin{split}
E\left(\tilde{u},t=-\frac{e^{-\tau}}{\sqrt{1-\alpha^2}}\right):=&\int_{\alpha t}^{\infty}\int_{\RR^3}\mid\nabla_{t,{\mathbf x},z}\tilde{u}(t,\mathbf{x},z)\mid^2z^{-3}d\mathbf{x}dz\\
=&-\int_{\Gamma_{\tau}}\left(t\mid\nabla_{t,{\mathbf x},z}u\mid^2+2\partial_tu\partial_zu\right)\frac{z^{-3}}{\sqrt{t^2+z^2}}d\Gamma_{\tau}\\
\geq&\int_{\Gamma_{\tau}}\mid t\mid\mid\nabla_{\mathbf x}u\mid^2\frac{z^{-3}}{\sqrt{t^2+z^2}}d\Gamma_{\tau}=:E_p(u,\tau),
\end{split}
\end{equation*}
then we return to the $\tau,\rho$ coordinates :
\begin{equation*}
\begin{split}
E_p(u,\tau)&=\int_{\frac{\alpha e^{-\tau}}{\sqrt{1-\alpha^2}}}^{\infty}\int_{\RR^3}\left\vert\nabla_{\mathbf x}u\left(t=-\sqrt{z^2+e^{-2\tau}},{\mathbf x},z\right)\right\vert^2z^{-3}d{\mathbf x}dz\\
&=e^{2\tau}\int_0^{\rho_0}\int_{\RR^3}\mid\nabla_{\mathbf x}u(\tau,{\mathbf x},\rho)\mid^2\sinh\rho\cosh\rho d{\mathbf x}d\rho\\
&\geq e^{2\tau}\Vert\nabla_{\mathbf x}u(\tau,.)\Vert^2_{X^0}.
\end{split}
 \label{}
\end{equation*}
We deduce from Proposition \ref{propufi} and Theorem \ref{theofluc} that
\begin{equation*}
\begin{split}
e^{2\tau}\Vert\nabla_{\mathbf x}u(\tau,.)\Vert^2_{X^0}&\gtrsim e^{2\tau}\Vert\nabla_{\mathbf x}u_{\varphi}(\tau,.)\Vert^2_{X^0}\\
&\gtrsim e^{2\tau}\Vert\nabla_{\mathbf x}\phi\Vert^2_{X^0}\\
&\gtrsim t^{-2}\Vert\nabla_{\mathbf x}\phi\Vert^2_{L^2(\RR^3)}.
\end{split}
 \label{}
\end{equation*}
That achieves the proof of the Theorem.\\
\fin

We conclude that the energy of $\tilde{u}(t=0,.)$ is infinite, hence the finite energy spaces used in \cite{braneg} cannot be used if we want to continue the solution in $[0,\infty[_t\times\RR^3_{\mathbf x}\times]0,\infty[_z$. Beyond the Cauchy horizon $\left\{(t,\mathbf{x},z)\in\RR^+\times\RR^3\times\RR^+,\;\;t=z\right\}$, it would be natural to consider a boundary constraint on the time-like conformal infinity $z=0$ in the new functional frameworks introduced in \cite{supersing} or \cite{warnick}.


\section{Acknowledgments}
This research was partly supported by the ANR funding ANR-12-BS01-012-01.


\end{document}